\documentclass[fleqn,usenatbib]{mnras} 

\usepackage[T1]{fontenc}

\DeclareRobustCommand{\VAN}[3]{#2}
\let\VANthebibliography\thebibliography
\def\thebibliography{\DeclareRobustCommand{\VAN}[3]{##3}\VANthebibliography}

\DeclareRobustCommand{\VANDER}[3]{#2}
\DeclareRobustCommand{\VON}[3]{#2}
\DeclareRobustCommand{\DE}[3]{#2}
\DeclareRobustCommand{\DELA}[3]{#2}

\usepackage{graphicx}	
\usepackage{amsmath}	
\usepackage{amssymb}	
\usepackage{multicol}   
\usepackage{bm}		    
\usepackage{newtxtext} %
\usepackage{siunitx}    
\usepackage{gensymb}
\usepackage{orcidlink}
\usepackage{booktabs} 

\newcommand{\planck}{\textit{Planck}}
\newcommand{\litebird}{\emph{LiteBIRD}}
\newcommand{\dust}{$d$}
\newcommand{\sync}{$s$}
\newcommand{\nv}{\hat{\bf n}}
\newcommand{\cov}{\textrm{Cov}}
\newcommand{\sigmar}{$\sigma(r)$~}
\newcommand{\Alens}{$A_{\textrm{lens}}$~}
\newcommand{\rtensor}{$r$}
\newcommand{\ampd}{$A_d$~}

\newcommand{\covng}{$\Sigma_{\textrm{NG}}$}
\newcommand{\covg}{$\Sigma_{\textrm{G}}$}
\newcommand{\bb}{\emph{BB}}
\newcommand{\B}{\emph{B}}
\newcommand{\DF}{\textsc{dustfilaments}}

\DeclareSIUnit{\yrs}{yrs.}

\defcitealias{BICEP_2021}{BK18}

\title[Non-Gaussian dust covariance]{\parbox[t]{\textwidth}{Impact of Galactic dust non-Gaussianity on searches for \emph{B}-modes from inflation}}

\author[Abril-Cabezas \textit{et al.}]{
\parbox[t]{\textwidth}{
Irene Abril-Cabezas \orcidlink{0000-0003-3230-4589},$^{1,2}${\thanks{E-mail: \href{mailto:ia404@cam.ac.uk}{ia404@cam.ac.uk} }}
Carlos Herv\'ias-Caimapo \orcidlink{0000-0002-4765-3426},$^{3}$
\mbox{Sebastian {von Hausegger} \orcidlink{0000-0002-6274-1424},$^{4,5}$}
Blake D. Sherwin$^{1,2}$ and
David Alonso \orcidlink{0000-0002-4598-9719}$^{5}$
}
\\ \\
\parbox[t]{\textwidth}{
$^{1}$Department of Applied Mathematics and Theoretical Physics, University of Cambridge, Cambridge CB3 0WA, United Kingdom\\
$^{2}$Kavli Institute for Cosmology, University of Cambridge, Cambridge CB3 0HA, United Kingdom\\
$^{3}$Instituto de Astrof\'isica and Centro de Astro-Ingenier\'ia, Facultad de F\'isica, Pontificia Universidad Cat\'olica de Chile, Av. Vicu\~na Mackenna 4860, 7820436, Macul, Santiago, Chile\\
$^{4}$Department of Physics, University of Oxford, Rudolf-Peierls Centre for Theoretical Physics, Parks Road OX1 3PU Oxford, United Kingdom\\
$^{5}$Department of Physics, University of Oxford, Denys Wilkinson Building, Keble Road, Oxford OX1 3RH, United Kingdom
}}

\date{Accepted 2023 November 13. Received 2023 November 8; in original form 2023 September 20.}

\pubyear{2024}
\volume{527}

\begin{document}
\label{firstpage}
\pagerange{5751--5766} 
\maketitle

\begin{abstract}
A key challenge in the search for primordial \B-modes is the presence of polarized Galactic foregrounds, especially thermal dust emission. Power-spectrum-based analysis methods generally assume the foregrounds to be Gaussian random fields when constructing a likelihood and computing the covariance matrix. In this paper, we investigate how non-Gaussianity in the dust field instead affects CMB and foreground parameter inference in the context of inflationary \B-mode searches, capturing this effect via modifications to the dust power-spectrum covariance matrix. For upcoming experiments such as the Simons Observatory, we find \emph{no} dependence of the tensor-to-scalar ratio uncertainty $\sigma(r)$ on the degree of dust non-Gaussianity or the nature of the dust covariance matrix. We provide an explanation of this result, noting that when frequency decorrelation is negligible, dust in mid-frequency channels is cleaned using high-frequency data in a way that is independent of the spatial statistics of dust. We show that our results hold also for non-zero levels of frequency decorrelation that are compatible with existing data. We find, however, that neglecting the impact of dust non-Gaussianity in the covariance matrix can lead to inaccuracies in goodness-of-fit metrics. Care must thus be taken when using such metrics to test \B-mode spectra and models, although we show that any such problems can be mitigated by using only cleaned spectrum combinations when computing goodness-of-fit statistics.
\end{abstract}




\section{Introduction}\label{sec:1}

The cosmic microwave background (CMB) was emitted when the Universe was approximately \num[group-separator = {,}]{380000} years old. The temperature and polarisation fluctuations present in the CMB have informed our understanding of the origin of the Universe. In particular, in the standard cosmological model these primordial density fluctuations were produced by quantum fluctuations in the inflationary early Universe. The simplest inflation models make a key prediction that has not yet been confirmed: the production of a primordial gravitational wave background \citep{Starobinski_1979, Abbott_1984}. This background would induce a unique \B-mode polarisation pattern in the CMB \citep{Kamion_1997, Seljak_1997}, which is an important signature of primordial tensor fluctuations. The amplitude of these fluctuations is parametrized by the tensor-to-scalar ratio \rtensor; the tightest constraint to date sets $r <  0.032$ at \SI{95}{\percent} confidence \citep{Tristram_2022}. Although $r$ could in principle be arbitrarily small, broad classes of well-motivated theories \citep[e.g.][]{Starobinski_1979} predict a value for $r$ that would be within the observable range of future CMB experiments, given current constraints on the scalar spectral index \citep[see][]{CMBS4_2016_Science}.

In this context, there are and will be multiple CMB experiments pursuing the search for the faint primordial \B-mode signal, such as \planck~\citep{Planck_2020_mission}, the BICEP2 and Keck array telescope \citep{BICEPKeck}, CLASS \citep{CLASS_2014}, SPT-3G \citep{Benson_2014}, the POLARBEAR-2 and Simons Array experiments \citep{POLSimons}, the Simons Observatory \citep[SO,][]{SimonsObs_2019}, the CMB-S4 experiment \citep{Abitbol_2017,Barron_2022}, and \litebird~\citep{Hazumi_2019, LiteBIRD_mission}. In particular, SO will deploy a number of Small Aperture Telescopes (SATs) targeted at measuring large-scale \B-modes to constrain \rtensor~at a level of \mbox{$\sigma(r = 0) \lesssim 0.003$} \citep{Wolz_2023}, using a effective sky fraction $f_{\textrm{sky}} \sim 0.1$ with a white noise level of around $\SI{2}{\mu\kelvin}$-arcmin. 

Unfortunately, the observed \B-mode signal is contaminated by Galactic foreground emission. Polarized synchrotron emission sourced by the motion of charged particles in the Galactic magnetic field (GMF) overwhelms the \B-mode signal at low ($\nu \lesssim \SI{70}{\giga\hertz}$) frequencies \citep[see e.g.][]{Page_2007}. In addition to this, thermal dust emission from our own Galaxy is dominant at frequencies $\nu \gtrsim \SI{70}{\giga\hertz}$ \citep[see e.g.][]{Plack_2016_X}. Dust grains, which are aligned with the GMF, are heated by interstellar radiation and subsequently re-emit partially polarized light that traces the GMF structure \citep{Draine_2009}. Dust is the most important foreground at most frequencies relevant for \B-mode searches.

As a consequence, exquisite control over the foreground polarisation signal is key to measuring primordial \B-modes. Several component separation methods have been developed to disentangle the primordial signal from foreground contamination. They can be broadly classified into three groups: methods working in harmonic space \citep[e.g.][]{Cardoso_2008,Dunkley_2013, BK14, BK15, Choi_2020, BICEP_2021, Mangilli_2021, Azzoni_2021}, map-based methods \citep[e.g.][]{Eriksen_2006, Gratton_2008,  Dunkley_2009, Efstathiou_2009, Remazeilles_2011, Stompor_2016, Chluba_2017, Hensley2018,  deBelsunce_2022, McBride_2023, Vacher_2023a} and hybrid methods \citep[e.g.][]{Fantaye_2011, Azzoni_2023}. They all require \B-mode polarisation measurements at a set of different frequencies to perform the cleaning. 

In this paper, we focus on the impact of Galactic dust non-Gaussianity on $C_\ell$-based methods.  These  methods perform a likelihood analysis of the power-spectrum signal, which includes all \bb~auto- and cross-power spectra at a set of frequencies. Crucially, they require an estimation of the power-spectrum covariance matrix. At this stage, most works usually assume that all sky components can be described by statistically isotropic Gaussian random fields at the map level. In the full sky, and assuming statistical isotropy, this would give rise to a purely diagonal covariance matrix, directly related to the power spectrum of the different components. Off-diagonal elements coupling different power-spectrum multipoles would then only be caused by masking and filtering of the data \citep{Knox_1997, Efstathiou_2004}. In practice, however, Galactic foregrounds are both statistically anisotropic and exhibit rather non-Gaussian features, both in temperature~\citep[e.g.][]{Ben-David:2015fia,Jung:2018rgf} and in polarisation~\cite[e.g.][]{vonHausegger:2018tjq,Coulton:2019bnz}, which inevitably leads to couplings between different scales. Given the magnetised and turbulent nature of the diffuse interstellar medium (ISM), which gives rise to highly non-linear motions in the ISM fluid, these couplings are indeed expected. At the map-level, they give rise to structures such as filaments. At the power-spectrum level, they introduce couplings between different $\ell$-bins in the power-spectrum covariance, which itself is no longer completely determined by just the multi-component power spectra. Although the problem of foreground non-Gaussianity has been briefly discussed in the literature \citep[e.g.][hereafter \citetalias{BICEP_2021}]{BICEP_2021}, an in-depth study on how dust non-Gaussianities affect the power-spectrum covariance, and the full impact this has on CMB and foreground parameter inference is still outstanding.

Traditionally, component separation studies have made use of simulations based on observed sky templates. Most of these templates rely on polarisation measurements that lack sensitivity on small scales. This issue is usually circumvented by filtering out these scales, and filling them in with Gaussian realizations of a given target power spectrum that follows the trend measured on large scales \citep[see e.g.][]{Hervias_2016, Thorne_2017}. As a result, works using these models do not account correctly for non-Gaussianity on small scales. In the last few years, several groups have produced foreground simulations including non-Gaussianities on small scales. These include magnetohydrodynamic simulations \citep[e.g.][]{Kritsuk_2018,Kim_2019}, phenomenological models based on temperature observations using a superposition of Galactic dust layers \citep[e.g.][]{Vansyngel_2017, Martinez_2018}, models built from wavelet scattering transform statistics \citep{Allys_2019, Regaldo_2020}, dust models based on neutral hydrogen \citep{Clark_2019}, and filament-based ones \citep{huffenberger_2020, DF_2022}. Similar studies have also been carried out in the context of Galactic synchrotron \citep[e.g.][]{Wang_2020, Martire_2022, Martire_2023}. \citetalias{BICEP_2021} explored some of these non-Gaussian foreground models. They used single dust realizations from \cite{Kritsuk_2018, Martinez_2018, Vansyngel_2017}, which they added to their modelled signal. However, with the analysis of only a limited number of realizations, it is unclear how reliably and precisely one can quantify the effect that foreground non-Gaussianity has on the $C_\ell$ covariance matrix and on the inference of \rtensor.

The paper is organized as follows. We present the general methodology of $C_\ell$-based analyses in Section \ref{sec:2}, with a particular focus on how to incorporate dust non-Gaussianity into the covariance matrix (\ref{sec:2-cov}). In Section \ref{sec:3}, we quantify the impact that dust non-Gaussianity has on parameter constraints, including in the presence of frequency decorrelation. We study the effect on parameter biases and variances, as well as goodness-of-fit metrics. We summarise our results and conclude in Section \ref{sec:4}.

\section{Formalism}\label{sec:2}

\subsection{Parameter inference}\label{sec:2-inference}

We focus on multi-frequency power-spectrum-based analyses. Here, the data vector is composed of all \bb~auto- and cross-power spectra between all frequency maps. We use Monte Carlo Bayesian analysis, as implemented in \textsc{emcee} \citep{emcee_2013} to obtain posterior distributions of our sky model parameters (Section \ref{sec:2-skymodel}) given the mock data vector we generated (Section \ref{sec:2-mock}). This analysis pipeline has already been used in works such as \cite{Azzoni_2021, Wolz_2023}.

Often, $C_\ell$-based analyses constraining primordial \B-modes assume a Gaussian likelihood within the Bayesian analysis. This implicitly assumes that each sky component can be summarised by its power spectrum and its covariance (ignoring measurement  noise and systematics). While this is certainly the case for the CMB, one also makes these assumptions for the foregrounds because each bandpower is an average over approximately $\Delta\ell \times (2\ell_{\textrm{eff}} + 1)$ modes, (neglecting correlations due to, e.g., masking), where $\ell_{\textrm{eff}}$ is the effective $\ell$ value at each bin of size $\Delta\ell$. Thanks to the high number of modes, the central limit theorem applies and one can assume that each $C_\ell$ follows a Gaussian distribution. However, this bandpower averaging does not remove the impact of non-Gaussian foregrounds, as the bandpower covariance matrix, with elements $\textrm{Cov}(C_\ell, C_{\ell^\prime})$, includes contributions from the connected trispectrum of the non-Gaussian foregrounds\footnote{As mentioned in Section \ref{sec:1}, when only a fraction of the sky is used, $f_{\textrm{sky}}< 1$, off-diagonal elements in the covariance matrix will also be present due to the mask.}.

We assess the impact of the non-Gaussian dust foreground by comparing two separate runs of the parameter inference. For these we use two different models for the covariance matrix, keeping the data vector and the sky model fixed. In the first case, we assume a Gaussian dust foreground. In the second case, we describe the anisotropic and non-Gaussian nature of the dust sky by modelling the covariance matrix accordingly. We denote these covariances as \covg~and \covng, respectively.  In Section \ref{sec:2-cov}, we describe how they are computed. 

We run our analyses with the \citet[][HL]{Hamimeche_2008} likelihood. It accounts for the non-Gaussian form of the likelihood on large scales, where there are fewer modes to average over.  On the scales we are interested in ($30 \leq \ell \leq 300$), both the Gaussian and HL likelihoods yield the same results \citep{Wolz_2023}. We explore the posterior distribution by Markov-Chain Monte Carlo sampling using 128 walkers and \SI[print-unity-mantissa=false]{1e4} steps, and compute the integrated autocorrelation time \citep{Goodman_2010} to assess convergence and ensure we have a sufficient number of independent samples. We remove the initial \SI{25}{\percent} of the chains (``burn-in'') to perform inferences. We present our results in Section \ref{sec:3}. 

\subsection{Sky model}\label{sec:2-skymodel}

To create the mock data vector in this work, we use the same sky model described in \cite{Abitbol_2021}, which follows the fiducial BICEP2/Keck model \citep[\citetalias{BICEP_2021};][]{BK15, BK14}. We assume three components $c$ for the sky signal: dust (\dust), synchrotron (\sync), and CMB. The foregrounds (\dust, \sync) auto-spectra are parametrized as a power-law:
\begin{equation}
    \frac{\ell(\ell + 1)}{2\pi}C_\ell^{cc} = A_c \left(\frac{\ell}{\ell_0}\right)^{\alpha_c}, \quad c = \{d,s\},
\end{equation}
where $A_c$ is the amplitude, $\alpha_c$ the tilt, and $\ell_0 = \SI{80}{}$ the pivot frequency. The model also allows for a cross-correlation between dust and synchrotron, 
\begin{equation}
    C_{\ell}^{ds} = \epsilon_{ds} \sqrt{C_{\ell}^{dd} C_{\ell}^{ss}},
\end{equation}
via a scale-independent parameter $\epsilon_{\textrm{\dust\sync}}$. The CMB \B-mode power spectrum is modelled as
\begin{equation}
        C_\ell^{\textrm{CMB}} = A_{\textrm{lens}}C_\ell^{\textrm{lens}} + r~C_\ell^{\textrm{tens}, r=1},
\end{equation}
where $C_\ell^{\textrm{lens}}$ and $C_\ell^{\textrm{tens}, r=1}$ are theoretical predictions obtained with \textsc{camb} \citep{CAMB_2011} for the \B-mode power spectrum produced by gravitational lensing and primordial tensor perturbations with $r=1$, respectively. Treating $A_{\rm lens}$ as a free parameter allows us to consider the need to marginalise over uncertainties in the residual lensing \B-modes if delensing is present ($A_{\rm lens}=1$ recovers the case without delensing), which will be of importance to improve $r$ constraints \citep[see e.g.][]{Seljak_2004, Sherwin_2015, Namikawa_2022}.

The multi-frequency power spectrum is then given by
\begin{equation}\label{eq:mfcl}
    C_\ell^{\nu\nu^\prime} = \sum_{c,c^\prime} S_c^{\nu}S_{c^\prime}^{\nu^\prime} C_\ell^{cc^\prime},
\end{equation}
where $S_c^{\nu}$ is the spectral energy distribution (SED) of component $c$ at frequency $\nu$ in thermodynamic temperature units. We model the SEDs in the same manner as \cite{Abitbol_2021}:

\begin{itemize}
    \item {\bf CMB.} For the CMB, the SED is a constant by definition:
    \begin{equation}
            S_\nu^{\rm{CMB}} = 1.
    \end{equation}
    \item {\bf Synchrotron.} We model the synchrotron spectrum as a simple power-law \citep[see e.g.][]{Planck_2020_diffuse}:
    \begin{equation}
        S_\nu^s = g(x_{\nu_0^s})\left(\frac{\nu}{\nu_0^s}\right)^{\beta_s},
    \end{equation}
    where $\beta_s$ is the synchrotron spectral index and $\nu_0^s = \SI{23}{\giga\hertz}$ is the pivot frequency. $g(x_\nu)$ is the conversion factor between Rayleigh-Jeans brightness temperature units (commonly used for foregrounds) and  thermodynamic temperature units \citep{Planck_2013_units}:
    \begin{equation}
            g(x_\nu) = e^x \left(\frac{x}{e^x - 1}\right)^2, \quad x = \frac{h\nu}{k \Theta_{\rm{CMB}}},
    \end{equation}
    where $\Theta_{\rm{CMB}} = \SI{2.7255}{\kelvin}$ is the CMB temperature \citep{Fixsen_2009}. 
        \item {\bf Dust.} Thermal dust emission is characterised by a modified black-body spectrum \citep[see e.g.][]{Planck_2020_diffuse}:
    \begin{equation}\label{eq:sed_dust}
        S_\nu^{d} = {g(x_{\nu_0^d})}\left( \frac{\nu}{\nu_0^d}\right)^{\beta_d} \frac{B_\nu(\Theta_d)}{B_{\nu_0^d}(\Theta_d)},
    \end{equation}
    where $\beta_d$ is the dust spectral index, $\nu_0^{d} = \SI{353}{\giga\hertz}$ is the pivot frequency, and $B_\nu(\Theta)$ is the Planck black-body spectrum, which expresses the spectral radiance of a black-body for frequency $\nu$ at temperature $\Theta$:
    \begin{equation}
        B_\nu(\Theta) = \frac{2h\nu^3}{c^2}\left[\exp{\left(\frac{h\nu}{k\Theta}\right) - 1}\right]^{-1},
    \end{equation}
    and where $k$ and $h$ are the Boltzmann and Planck constants, respectively. The priors for all model parameters presented hereto are based on table 1 of \cite{Azzoni_2021}, which we reproduce in Table \ref{tab:priors} for completeness. 
\end{itemize}

\begin{table}
    \centering
    \caption{Priors used in the analysis, based on those used in \protect\citet{Azzoni_2021}. The power-spectrum amplitude for tensor modes $r$ is allowed to be negative to detect possible biases.}
    \begin{tabular}{lcc}
    \toprule
    Parameter & Prior & Bounds  \\ \midrule
    $r$ &  Top-hat & $[ -1,1 ]$\\ 
    $A_{\rm{lens}}$& Top-hat& $[ 0,2 ]$\\ 
    $A_d$ &Top-hat & $[ 0, \infty)$\\
    $\alpha_d$ & Top-hat&$[ -1,0]$ \\
    $\beta_d$ & Gaussian &$1.59\pm 0.50 $ \\
    $A_s$ & Top-hat& $[0, \infty)$\\
    $\alpha_s$ & Top-hat& $[-1,0 ]$\\
    $\beta_s$ & Gaussian & $  -3.0\pm 0.6$ \\
    $\epsilon_{ds}$ &Top-hat & $[-1,1 ]$\\
    $B_d$ & Top-hat& $[ -10, 10]$\\
    $\gamma_d$ &Top-hat &$[ -6,-2]$ \\
    $B_s$ &Top-hat & $[-10, 10 ]$\\
    $\gamma_s$ &Top-hat & $[ -6,-2]$\\
    $\Delta_d$ & Top-hat& $[0.9, 1.1]$ \\
    \bottomrule
    \end{tabular}
    \label{tab:priors}
\end{table}
\begin{figure*}
     \centering
     \includegraphics[width = 0.9\textwidth]{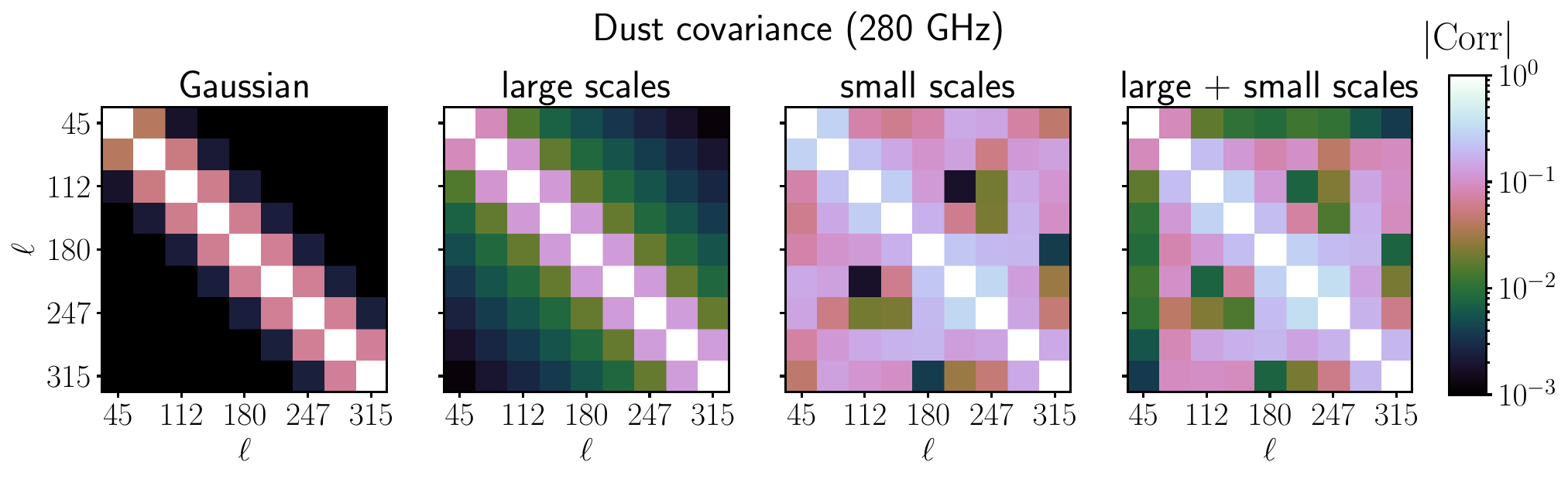}
     \caption{Correlation of the dust power spectrum covariance at \SI{280}{\giga\hertz} for different dust models in the SO patch. The $\ell$ range spans $\ell \in (30,300)$, binned with $\Delta \ell = 30$, and is the same across all panels. The first panel corresponds to the covariance matrix of our initial Gaussian dust field, \covg, which already incorporates masking effects: the covariance is essentially diagonal with the exception of small off-diagonal terms close to the diagonal that appear due to the mask. The second panel corresponds to a Gaussian dust field modulated by a large-scale template (Section \ref{sec:2-large}). One can see that the large-scale modulation moves power from the diagonal to the off-diagonal terms, thus further coupling neighbouring $\ell$-bins. The third panel corresponds to the covariance of a filamentary small-scale dust population obtained from 200 realizations of the \DF~\protect\citep{DF_2022} model. The net effect of this population is to couple small scales to large scales. The last panel refers to the covariance matrix \covng, which contains both large and small scale effects. We use it to capture the full anisotropic non-Gaussian nature of the dust field.}
     \label{fig:corr_cov_comparison}
 \end{figure*}
The simple frequency dependence of equation \eqref{eq:mfcl} assumes that all components are perfectly correlated across frequencies. However, we expect Galactic foregrounds to present some level of frequency decorrelation, caused by variations in their spectral energy distribution across the sky and along the line of sight \citep[see e.g.][]{Pelgrims_2021, Ritacco_2023}. To parametrize this decorrelation we will make use of a simple scale-independent dust decorrelation parameter, $\Delta_d$, quantifying the map-level correlation coefficient between dust at $353$ and \SI{217}{\giga\hertz} \citep{planck_int_XXX}. Specifically, it is defined as the ratio of the cross-spectrum between those two maps and the geometric mean of the corresponding auto-spectra \citep[see equation 1 in][]{Planck_2017_spatial}. Even though the physically meaningful range for $\Delta_d$ is $\Delta_d \leq 1$ \citep{BK15}, we allow for $\Delta_d$ values greater than $1$ to check for possible biases. We will also assess the impact of frequency decorrelation through the minimal moment expansion method of \cite{Azzoni_2021}, which is based on the ``moments-based''  expansion \citep{Chluba_2017} and effectively generalises of the simple $\Delta_d$ parametrization, modelling the spatial fluctuations of the foreground spectral indices, $\delta\beta_c$, as Gaussian random fields with a power spectrum of the form
\begin{equation}\label{eq:moments}
    C_\ell^{\beta_c} = B_c \left(\frac{\ell}{\ell_0}\right)^{\gamma_c}, \quad c = \{d,s\},
\end{equation}
and associated frequency dependence \citep{Azzoni_2021}
\begin{equation}
\tilde{S}_\nu^c\left( \delta \beta_c\right) = \log{\left( \frac{\nu}{\nu_0^c}\right)} S_\nu^c, \quad c = \{d,s\},
\end{equation}
Even though $B_s, B_d$ correspond to auto-spectra amplitudes \citep[see e.g.][]{Mangilli_2021, Vacher_2022}, we do not enforce positivity in their priors. This is to avoid a bias in the marginalised posterior distribution of $r$, which is found when data with no spectral index variations are analysed with the moments expansion and a positive prior \mbox{$B_c \geq 0$} \citep{Azzoni_2021}.  In summary, our baseline sky model has 9 free parameters.  The number of parameters is extended to 10 as we include decorrelation in the dust component, or to 13 for the moments expansion in both dust and synchrotron.

\subsection{Mock data}\label{sec:2-mock}

We generate mock multi-frequency power-spectrum data on which to run our analysis by evaluating the baseline model we described in the previous section (\ref{sec:2-skymodel}) with the following parameter values. For the CMB contribution, we assume no primordial \B-modes, $r = 0$, and assume no delensing, $A_{\textrm{lens}} = 1$, in the first instance. For the foreground power spectra, we use ${A_d = \SI{28}{\mu\kelvin\squared}}$, ${\alpha_d = \SI{-0.16}{}}$, ${A_s=\SI{2.0}{\mu\kelvin\squared}}$, and ${\alpha_s = \SI{-0.60}{}}$, and assume no dust-synchrotron correlation (${\epsilon_{ds} = 0}$). These values correspond to the best-fitting amplitudes found in \cite{Azzoni_2021} from a suite of \textsc{pysm} \citep{Thorne_2017} foreground simulations within the SO footprint\footnote{See, however, nuances on $A_d$ and $\alpha_d$ in Section~\ref{sec:2-small}.}. For the SED parameters, we use the best-fitting parameters from \planck: ${\Theta_d = \SI{19.6}{\kelvin}}$, $\beta_d = \SI{1.59}{}$ \citep{Planck_2015_dust} and $\beta_s = \SI{-3.00}{}$ \citep{Dunkley_2009, Planck_2016_low}. 

We take the Simons Observatory experiment, as described in \cite{SO_2019}, as our baseline instrument model. We generate multi-frequency power spectra using the sky model above assuming six SO frequency bandpasses, centred around \SI{27}{} and \SI{39}{\giga\hertz} (low-frequency bands, LF), \SI{93}{} and \SI{145}{\giga\hertz} (mid-frequency bands, MF), and \SI{225}{} and \SI{280}{\giga\hertz} (ultra-high-frequency bands, UHF). This leads to a total of 21 unique multi-frequency power spectra $C_\ell^{\nu\nu^\prime}$.

To incorporate the contribution from instrumental noise to the covariance matrix, we include the noise auto-spectrum at each frequency, assuming no correlation of the noise between different frequencies. For this we follow \cite{SO_2019} and model the SO SATs' noise spectrum as:
\begin{equation}\label{eq:noiseatm}
  {N}^{\nu\nu'}_\ell = \delta_{\nu\nu'}{N}^\nu_{\textrm{white}} \left[ 1 + \left(\frac{\ell}{\ell_{\textrm{knee}}}\right)^{\alpha^\nu_{\textrm{knee}}}\right],
\end{equation}
where the first term accounts for white noise and the second for red $1/f$ noise, most prominent on large scales, and caused by atmospheric contributions and timestream filtering. This noise model is rather simplistic, and does not incorporate features associated with the specific scanning strategy, such as inhomogeneity or anisotropy. Although \cite{Wolz_2023} find that noise inhomogeneity may lead to an increase in $\sigma(r)$ of up to $\sim \mathcal{O}(\SI{30}{\percent})$, this level of precision is sufficient for the purposes of quantifying the impact of foreground non-Gaussianity. We use values of ${N}^\nu_{\rm white}$ and $\alpha^\nu_{\rm knee}$ corresponding to the ``goal-optimistic'' scenario, specified in tables 1 and 2 of \cite{SO_2019}. This corresponds to the most optimistic noise model envisaged by SO. 

We apply our results to \bb-power-spectrum measurements on the SO patch, which we take to be our baseline analysis. We use the SAT footprint released in \cite{SO_2019} as our sky mask. Its edges are apodized on \ang{5} scales using the ``C1'' apodization method \citep{Grain_2009} as implemented in \textsc{namaster}\footnote{\url{https://github.com/LSSTDESC/NaMaster/}} \citep{namaster}, giving an effective sky fraction of $f_{\textrm{sky}} = 0.1$. We only use data in the range $30 \leq \ell \leq 300$, binned into $\Delta\ell = 30$ (9 bins in total). We choose a somewhat broad $\ell$-binning to ensure that the covariance converges given the number of dust filament realizations we have (see Section \ref{sec:2-small}), but we check that our results are robust against a finer $\Delta\ell=15$ binning.

\subsection{Dust non-Gaussian covariance}\label{sec:2-cov}

As we describe in Section \ref{sec:2-inference}, we capture the effect of the non-Gaussian dust foreground in the covariance matrix \covng.  This is then compared to the results obtained assuming a Gaussian dust foreground as per \covg. We first sketch how we compute the Gaussian covariance matrix \covg. Detailed calculations of all results obtained in this section are described in Appendix \ref{sec:ap1}.

We start with a Gaussian field $\delta(\nv)$\footnote{In this work, we only work with the \B-mode component, which we treat as a spin-0 scalar field.}, which is modulated by the sky mask $w(\nv)$:
\begin{equation}\label{eq:dG}
    d_{\textrm{G}}(\nv) = w(\nv)\delta(\nv),
\end{equation}
and define the corresponding pseudo-$C_\ell$, $\tilde{C}^{d_{\textrm{G}}}$, for which one has
\begin{equation}\label{eq:modec}
\langle\tilde{C}^{d_\textrm{G}}_\ell\rangle =\sum_{\ell^\prime}M^{w}_{\ell\ell^\prime}\,C_{\ell'}^\delta,
\end{equation}
where $C^\delta_\ell$ is the power spectrum of $\delta(\nv)$, and we have defined the $C_\ell$-level mode-coupling matrix $M^{w}_{\ell\ell^\prime}$, which can be computed from the sky mask alone \citep{Hivon_2002,namaster}. From equation \eqref{eq:modec}, one can see that the effect of the sky mask is to couple different $\ell$-modes. 

We calculate the covariance of the power spectrum of the Gaussian field via its 4-point function. We assume that the sky mask is smooth, the mode-coupling coefficients are peaked around $\ell = \ell^\prime$ and that $C_\ell^{\delta}$ is slowly varying. The final expression is:
\begin{equation}\label{eq:covG}
     {\rm Cov}({C}^{\delta}_{\ell_1},{C}^{\delta}_{\ell_2}) =\frac{2(\bar{C}^\delta_{\ell_{12}})^2}{\langle w^2 \rangle^2_\Omega\left(2\ell_2+1\right)}M^{w^2}_{\ell_1\ell_2},
\end{equation}
where $\bar{C}^\delta_{\ell_{12}}$ denotes the geometric mean of $C_{\ell_1}^\delta$ and $C_{\ell_2}^{\delta}$, and $\langle\cdots\rangle_\Omega$ denotes sky averaging. $M^{w^2}_{\ell_1\ell_2}$ is now the mode-coupling matrix for the sky mask squared, $w^2(\nv)$, which we compute analytically with \textsc{namaster} \citep{namaster}.  Under the given approximations, the $\langle w^2 \rangle_\Omega$ term serves as the conversion factor between $\tilde{C}^{d_\textrm{G}}_\ell$ and $C_\ell^\delta$.  Equation \eqref{eq:covG} is essentially the Knox formula \citep{Knox_1995, Knox_1997} modified to properly include mask-coupling effects \citep{Efstathiou_2004,Garcia_2019}. We show in the first panel of Figure \ref{fig:corr_cov_comparison} the correlation matrix for the power-spectrum covariance of a Gaussian dust field at \SI{280}{\giga\hertz} computed in this way. The covariance \covg~is essentially diagonal with the exception of small coupling terms close to the diagonal that appear due to the mask.

However, as we already discussed, the nature of foregrounds is anisotropic and non-Gaussian, and quantifying the impact of this on parameter inference is the main goal of this paper. Previous analyses \citep[e.g.][]{Azzoni_2023} have assumed Gaussian foregrounds or have only briefly discussed the question of how non-Gaussian foregrounds impact \rtensor~inference \citepalias{BICEP_2021}. Focusing on dust, we compare how the covariance matrix of dust changes as we introduce non-Gaussianity to an initially Gaussian isotropic dust field in two regimes:
\begin{itemize}
    \item {\bf Large-scale modulation.} We introduce non-Gaussianity by modulating the Gaussian dust field by a large-scale template, which we take to be the \planck~\SI{353}{\giga\hertz} temperature map smoothed on small scales. This model was already proposed in \cite{Mak_2017}. The net effect of the modulation is to couple neighbouring bins, thus moving power from the diagonal to the nearby off-diagonal elements of the covariance.
    \item {\bf Small-scale dust filaments.} We use dust simulations based on filaments \citep[\DF,][]{DF_2022} to estimate the covariance on small scales. These structures affect all scales and therefore introduce dominant off-diagonal elements away from the diagonal. 
\end{itemize}

\subsubsection{Large-scale modulating template}\label{sec:2-large}

\begin{figure}
    \centering \includegraphics[width=\columnwidth]{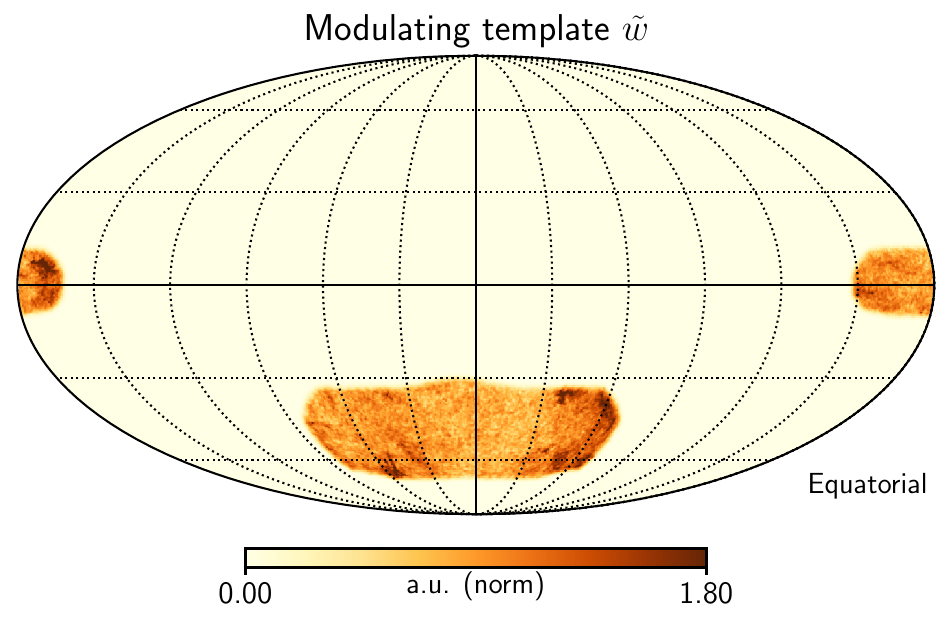}
    \caption{Large-scale modulating template in normalised arbitrary units (a.u.) for the SAT region of the SO experiment. It is the product of a uniform SO patch mask (apodized on \ang{5} scales around the edges) with the large-scale field. The large-scale field corresponds to the Planck \SI{353}{\giga\hertz} temperature map, smoothed on \ang{0.4} ($\ell \approx 400$) scales. We take this map to modulate the Gaussian dust field, introducing anisotropic features which are reflected as couplings between neighbouring $\ell$-bins in the power-spectrum covariance.}
\label{fig:template_353_so}
\end{figure}

As found in \cite{Miville_2007}, the small-scale variance of the dust emission scales with the square of the large-scale intensity. This relation motivates a simple model to account for the statistically anisotropic nature of Galactic dust, as proposed in  \cite{Mak_2017}. In this model, we modify equation \eqref{eq:dG} so that the Gaussian isotropic dust field $\delta(\nv)$ is also modulated by a large-scale intensity field $t(\nv)$:
\begin{gather}
    d_{\textrm{NG}}(\nv) = w(\nv)t(\nv)\delta(\nv),
\end{gather}
which enforces the results found empirically. There are multiple ways to estimate the template $t(\nv)$ \citep{Delabrouille_2013, Remazeilles_2015, Hervias_2016, Thorne_2017, Mak_2017}. In this work, we take as modulating template the \SI{353}{\giga\hertz} temperature map from Planck \citep{Planck_2020_HFI}, smoothed on \ang{0.4} ($\ell \approx 400$) scales. This scale is chosen to be well above the map noise limit, which we estimate using the FFP10 simulations \citep{Planck_2020_HFI}. We run our analysis for different smoothing scale values in the range \ang{0.1} -- \ang{40.0} and check that our final results are robust against changes in the smoothing scale. The results we present in Section \ref{sec:3} correspond to a somewhat aggressive choice of smoothing scale (\ang{0.4}), where the non-Gaussian component is enhanced in comparison to the anisotropy that would be present in a map with stronger smoothing.

After smoothing, the template is normalized to ensure that both the Gaussian dust field and the modulated one have the same power spectrum (see Appendix \ref{sec:ap1-norm}). As a result, the non-Gaussianity only enters at the level of the power-spectrum covariance. We show the modulating template on the SO patch in Figure \ref{fig:template_353_so}.

In Appendix \ref{sec:ap1}, we derive how the dust power-spectrum covariance changes as we account for the modulating template. The idea is to capture the effect of $t(\nv)$ as producing an effective sky mask $\tilde{w}(\nv)\equiv w(\nv) t(\nv)$ and repeat the procedure used for the masked Gaussian field. The template normalization ensures that $\langle w^2 \rangle^2_\Omega = \langle \tilde{w}^2 \rangle^2_\Omega$. The new effective mask introduces new couplings between neighbouring $\ell$-bins. This is in turn reflected in the covariance matrix of the dust field $D(\nv) = t(\nv)\delta(\nv)$: 
\begin{equation}\label{eq:covNG}
     {\rm Cov}({C}^{D}_{\ell_1},{C}^{D}_{\ell_2}) =\frac{2(\bar{C}^\delta_{\ell_{12}})^2}{\langle w^2 \rangle^2_\Omega\left(2\ell_2+1\right)}M^{\tilde{w}^2}_{\ell_1\ell_2},
\end{equation}
where the mode-coupling matrix now captures both the effect of the mask and the template via the effective mask squared, $\tilde{w}^2(\nv)$. We show how it compares to the Gaussian case in the first two panels of Figure \ref{fig:corr_cov_comparison}. As can be seen in the second panel, the large-scale modulating template increases the coupling between neighbouring bins and introduces a larger variance with respect to the Gaussian case (first panel).

\subsubsection{Small scales}\label{sec:2-small}

\begin{figure*}
    \centering
    \includegraphics[width = 0.7\textwidth]{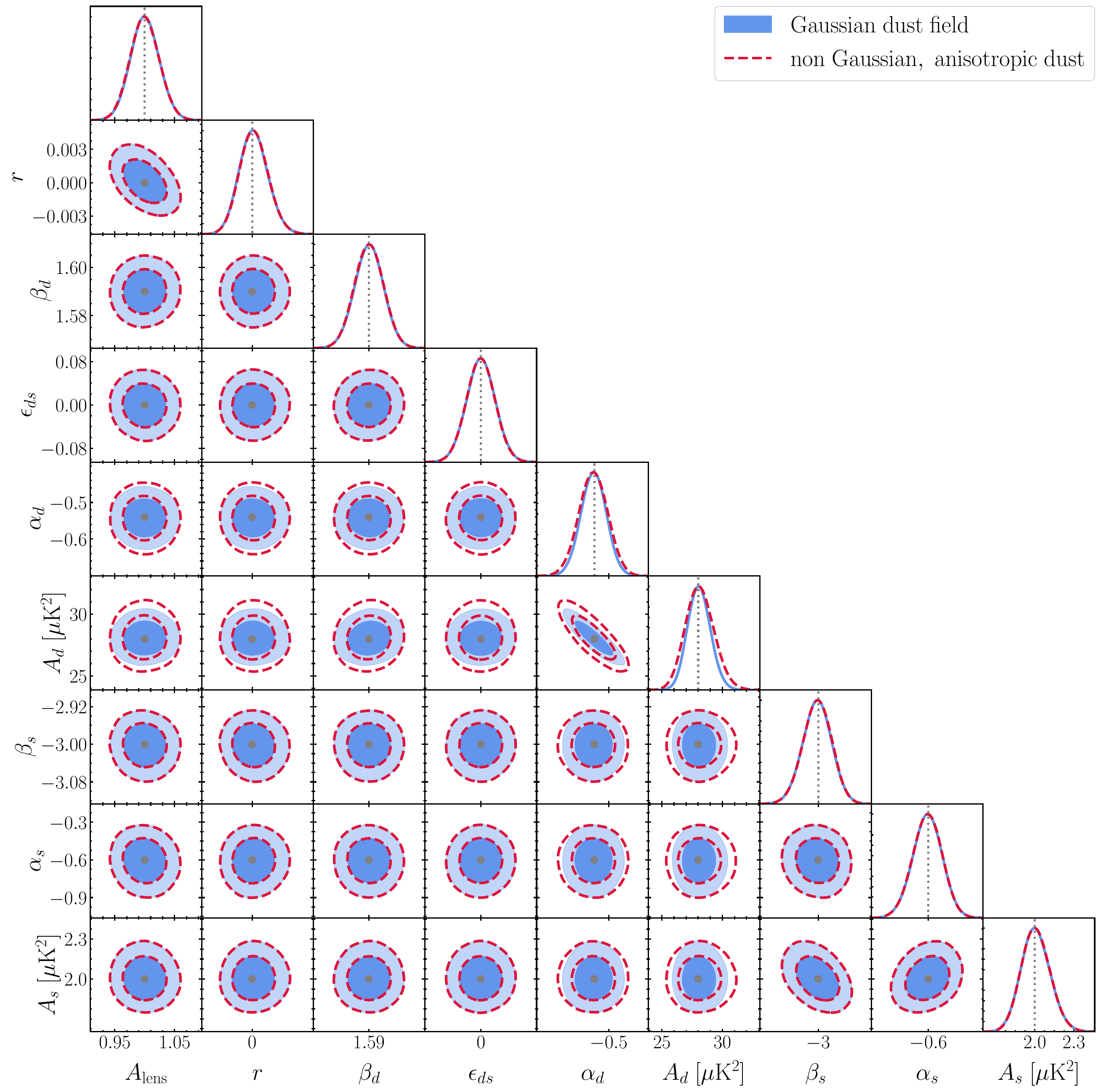}
    \caption{Comparison between posterior distributions obtained with a Gaussian and non-Gaussian dust covariance matrix (shown in filled solid blue and unfilled dashed red, respectively). The key result is that \sigmar is \emph{unaffected}, while the spatial dust parameter uncertainties are greater in the non-Gaussian case. When we use the covariance matrix capturing the non-Gaussian anisotropic nature of dust, we see an increase in $\sigma(\alpha_d)$ and $\sigma(A_d)$ by a factor $1.1$ and $1.3$, respectively. Therefore, using a Gaussian covariance to describe the statistics of dust leads to an underestimation of the dust spatial parameter uncertainties. True mock data values are shown as grey dotted lines. We recover unbiased parameter constraints in all cases. This is expected since non-Gaussian dust only affects the covariance matrix in our analysis. }    \label{fig:posterior_baseline}
\end{figure*}

The previous model assumes that the small-scale fluctuations are Gaussian distributed and modulated by a template. In practice, we know that this is a poor description for Galactic dust, which, for instance, exhibits a filamentary structure that cannot be reproduced from random Gaussian statistics. These filaments have been observed at millimetre wavelengths by \planck~\citep[e.g.][]{planck_int_XXXIII,planck_int_XXXVIII}, among others, and also found to be traced by Galactic neutral hydrogen emission \citep{Clark_2014,Clark_2015, Clark_2019, Halal_2023}. To improve the modelling of small-scale fluctuations, we use the dust filament model \DF\footnote{\url{https://github.com/ chervias/DustFilaments}} \citep{DF_2022}. Filaments and their misalignment with respect to the Galactic magnetic field can quantitatively reproduce the properties of Galactic dust as observed by \planck~at \SI{353}{\giga\hertz} \citep{planck_int_XXX,Planck_2020_dust}, such as the \textit{BB/EE} power ratio or the dust \bb-power-spectra power-law slope  $\alpha_d$ \citep{huffenberger_2020}. Based on this idea, \DF~constructs a full-sky simulation of the Galactic dust at millimetre frequencies by integrating the emission of millions of filaments in a predefined volume.

By construction, the filament population is calibrated to match the Large Region 71 (LR71) Galactic mask in \cite{Planck_2020_dust}, as described in \cite{DF_2022}. Since the polarization fraction changes depending on the portion of the sky considered, the \DF~model would produce a different amplitude of polarization signal if calculated in a different region than LR71, e.g. the SO patch \citep[see section~5 of][for an extended discussion of this point]{DF_2022}. Consequently, we re-scale the final \DF~power spectra to match our chosen value for $A_d$ in the SO patch. Moreover, the slope $\alpha_d$ in the simulations has the value found for the LR71 mask ($\alpha_d = -0.54$) by default. Our mock data therefore agree with this value, instead of the number previously quoted in Section \ref{sec:2-mock}.

The raw \DF~model does not reproduce the large-scale features in polarization. Before computing the covariance matrix from a set of 200 realizations, we employ a harmonic high-pass filter that removes the large-scale signal $\ell \lesssim 50$ from each map. The resulting dust power-spectrum covariance matrix on small scales is shown in the third panel of Figure \ref{fig:corr_cov_comparison}. This covariance is sourced only by the filamentary small-scale dust population, which induces important correlations between small and large scales. This effectively adds off-diagonal terms to the power-spectrum covariance and also leads to larger small-scale variance than predicted by the Gaussian covariance. We also check that the power spectra from which the covariance is derived follow a Gaussian distribution once they are binned in bandpowers of width $\Delta\ell = 30$. For this, we make histograms of the distribution of power spectra across the 200 simulations for the lowest and highest bandpower. We check that they closely follow a Gaussian distribution with a mean given by the bandpower mean and a standard deviation derived from the diagonal of the covariance matrix.

Since we have filtered out the large scales,  we cannot estimate the covariance matrix at low $\ell$ directly from the simulations. Consequently, we merge the covariance obtained from small scales with the one obtained in the previous section for large scales.  The procedure is described in Appendix \ref{sec:ap-sumcov}. It involves computing power spectra at every $\ell$ using the usual \texttt{anafast} \textsc{healpix}\footnote{\url{http://healpix.sf.net}} routine \citep{Healpix_2005}. We show the resulting correlation coefficient for the covariance matrix in the last panel of Figure \ref{fig:corr_cov_comparison}. This final covariance matrix \covng~is the one we use to represent the non-Gaussian anisotropic dust field.

\section{Results and discussion} \label{sec:3}

\subsection{Baseline analysis}\label{sec:3-baseline}

We show in Figure \ref{fig:posterior_baseline} the 2- and 1-dimensional marginalised posterior distributions for the 9 theoretical model parameters. We compare the posterior distributions obtained when the likelihood incorporates the covariance matrix of a Gaussian dust field (\covg, filled solid blue) with the contours obtained when assuming a non-Gaussian anisotropic dust field (\covng, unfilled dashed red). Note that the only difference between the two cases lies in the dust contribution to the covariance; the mock power-spectra remain the same in both runs regardless of the employed covariance, as discussed in Section \ref{sec:2-inference}.

Judging from this plot we can describe how dust non-Gaussianities impact $C_\ell$-based analyses on primordial \B-modes. We mark in dotted gray the values used to generate the mock data, from which one can see that we obtain unbiased constraints on all parameters. This is expected since non-Gaussian dust only affects the covariance matrix in our analysis. More importantly, the uncertainties of both \rtensor~and \Alens are insensitive to the anisotropic and non-Gaussian nature of the dust covariance matrix. We recover $\sigma(r) =\SI{1.3e-3}{}$, in agreement with previous forecasts focused on SO that used the same noise model \citep{SO_2019, Azzoni_2021, Wolz_2023}\footnote{The $\sigma(r)$ value we report is slightly lower than other works forecasting SO performance, for which they assume a higher white noise level \citep[\mbox{$\sigma(r) = \SI{1.7e-3}{}$},][]{Azzoni_2021} or inhomogeneous noise \citep[\mbox{$\sigma(r)=\SI{2.1e-3}{}$},][]{Wolz_2023}}. We note that this result is also consistent with the \citetalias{BICEP_2021} test, which also found no effect on their \rtensor~constraints when using dust simulations whose amplitude is modulated by degree-scale \bb~power measured on the \planck~\SI{353}{\giga\hertz} map. However, for the dust power-spectrum parameters \ampd and $\alpha_d$, we do see a difference in their posterior variance depending on the covariance matrix used. In the Gaussian case, we have $A_d = 28.08 \pm 0.93$, $\alpha_d = -0.542 \pm 0.035$, as opposed to $A_d = 28.1 \pm 1.2$, $\alpha_d = -0.543 \pm 0.040$ in the non-Gaussian case; the non-Gaussian uncertainties are larger by a factor $1.3$ and $1.1$, respectively. Therefore, using a Gaussian covariance to describe the statistics of dust leads to an underestimation of the dust spatial parameter uncertainties.

We also find the same qualitative behaviour in a more realistic case that includes delensing at the level expected for SO, $A_{\textrm{lens}} = 0.3$ \citep{Namikawa_2022}. Although the off-diagonal entries of the \B-mode power-spectrum covariance will become somewhat more relevant after delensing, the lensing \B-modes are typically small compared to the dust foreground power, so this result is not surprising. In this paper, we do not investigate the impact of dust non-Gaussianity in the delensing tracer itself, as the results will be dependent on which tracer is used and the bias arising from non-Gaussian dust in delensing has already been investigated in other works \citep{Beck_2020,Baleato_2022}.

\begin{figure}
    \centering
    \includegraphics[width = 0.9\columnwidth]{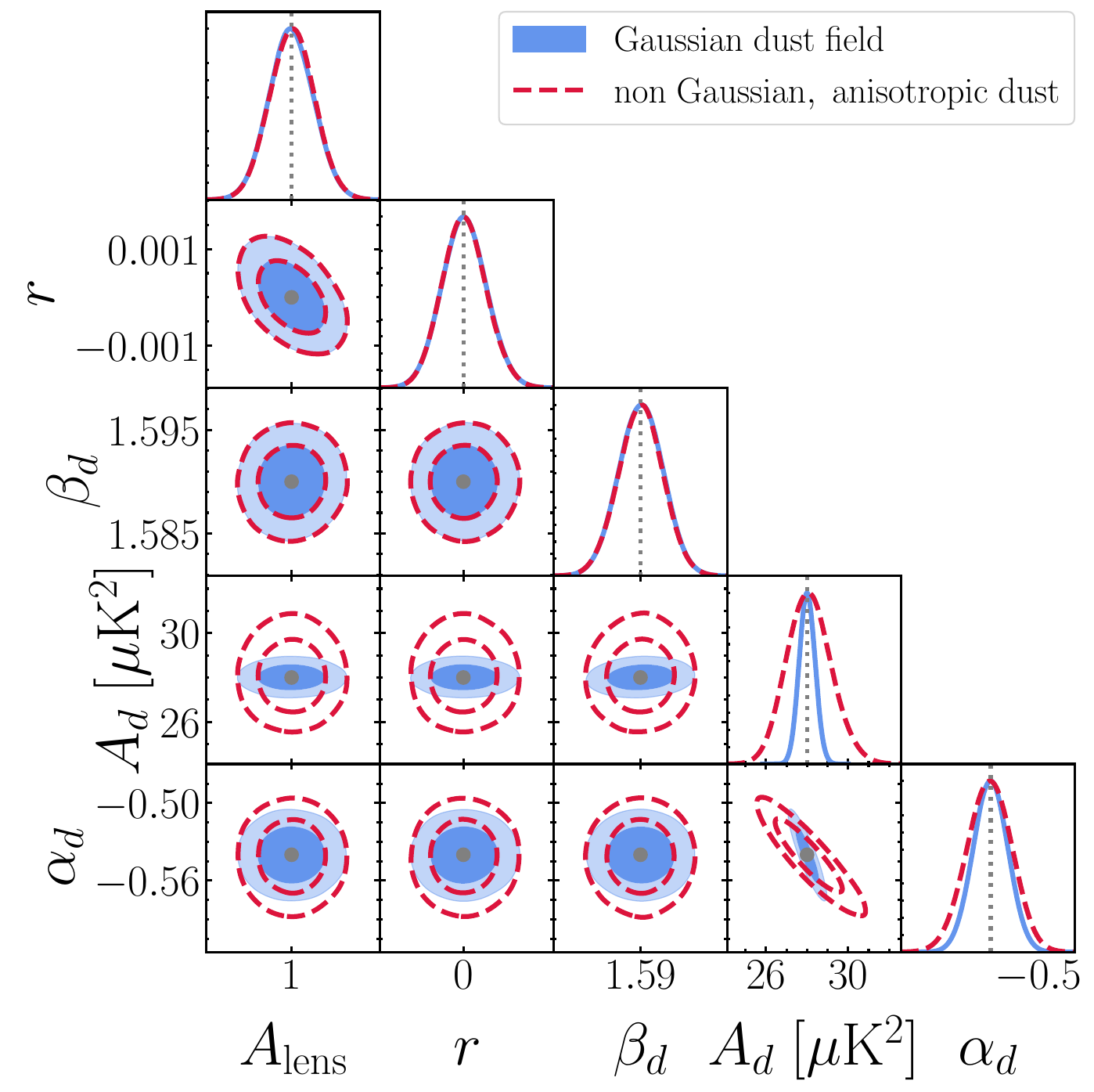} 
    \caption{Comparison between posterior distributions obtained with a Gaussian and non-Gaussian dust covariance matrix (shown in filled solid blue and unfilled dashed red, respectively). We only show the most relevant parameters included in the sky model. We employ a configuration similar to a \litebird-like experiment and artificially boost \covng~ to a level corresponding to a (very large) $\SI{10}{\percent}$ correlation matrix. We find the same results as our baseline analysis: we recover the same unbiased parameter constraints in both (\covg, \covng) cases, with the exception of the dust spatial parameters ($A_d$, $\alpha_d$), whose variance increases. The larger sky fraction and decrease in noise levels improves the error on $r$, $\sigma(r)=\SI{5.0e-4}{}$.}   \label{fig:posterior_crazy}
\end{figure}

We also check that this result holds for larger sky coverages ($f_{\textrm{sky}} \sim 0.6$) in a noise configuration without atmospheric noise (similar to a \litebird-like experiment). In this case, the increase in sky fraction and decrease in noise levels improves the error on \rtensor, $\sigma(r) = \SI{5.0e-4}{}$, for both \covg~and \covng. Remarkably, we observe the same behaviour in this configuration even when we make a large change to the dust covariance matrix, artificially boosting every off-diagonal dust covariance matrix element to a level corresponding to a (very large) \SI{10}{\percent} correlation matrix: the error on \rtensor~remains at $\sigma(r) = \SI{5.0e-4}{}$, whereas now the errors on the spatial dust parameters $A_d$ and $\alpha_d$ increase by a factor $2.8$ and $1.3$, respectively. We show the corresponding posteriors in Figure \ref{fig:posterior_crazy}.

We also explore whether dust non-Gaussianities (incorporated via the power-spectrum covariance matrix) introduce biases in the recovered parameter values. We generate \SI[print-unity-mantissa=false]{1e5} mock power-spectrum observations $\hat{C}_\ell$ by performing a multivariate Gaussian sampling of the non-Gaussian power-spectrum covariance matrix \covng. For each mock observation, we find the best-fitting sky model parameters. We find that the histogram of best-fitting values found for $r$ and the spatial dust parameters are the same regardless of whether we use \covng~or \covg~when evaluating the likelihood, and these are centred around the true model parameter values (Section \ref{sec:2-skymodel}). We thus expect dust non-Gaussianities to introduce no bias in our measurements.

We thus see that the presence of significant dust non-Gaussianity does not seem to affect the final constraints on $r$, and its impact is reduced to those parameters describing the spatial distribution of dust. Understanding this result is the goal of the next section.

\subsection{Theoretical model for observed behaviour}\label{sec:3-fisher}

To gain insight into the insensitivity of the constraints on $r$ to dust non-Gaussianity, we employ a simple Fisher forecast to estimate parameter constraints analytically. We start by writing down a simple model \citep{Efstathiou_2019} for an experiment which measures only two frequencies: $M$ (mid-frequency band, where the primordial CMB signal is present) and $H$ (high-frequency band, dominated by dust). Ignoring any other foregrounds and lensing for simplicity, we write the data vector at the power-spectrum level as: 
\begin{equation}\label{eq:model}
    \vec{\mu} = \begin{pmatrix}
        C_\ell^{MM}\\ C_\ell^{MH} \\ C_\ell^{HH}\end{pmatrix} = \begin{pmatrix}
            f^2 C_\ell^{dd} + N_\ell^{M} + r C_\ell^{BB} \\
            f C_\ell^{dd} \\
            C_\ell^{dd} + N_\ell^{H} 
    \end{pmatrix},
\end{equation}
where we have included a noise signal $N_\ell^\nu$ on each channel, which captures other terms we are neglecting, such as the lensing \B-mode power in the mid-frequency channel or the \B-mode tensor contribution in the high-frequency channel. This is a two-parameter model. $f$ controls the ratio of the dust SED between both frequency channels, while $r$ is the amplitude of the primordial \bb-power spectrum. As before, we invoke the central limit theorem and write down a Gaussian likelihood for which the Fisher matrix elements take the form \citep[see e.g.][]{Heavens_2016}:
\begin{equation}\label{eq:fisher}
    F_{\alpha\beta} = \vec{\mu}_{,\alpha}^{T}\,\Sigma^{-1} \vec{\mu}_{,\beta},
\end{equation}
where $\Sigma$ is the (diagonal) covariance matrix, $\mu$ is the mean of the data and where we have denoted partial derivatives with respect to model parameters as ``$_{,\alpha}$''. We have also ignored parameter dependences in the covariance matrix, as is appropriate \citep{Carron_2013}. 

Assuming Gaussian fields, we calculate the (diagonal) covariance using the Wick's theorem \citep[see e.g.][]{Allison_2015}:
\begin{equation}
    \textrm{Cov}(C_\ell^{\nu_1\nu_2}, C_\ell^{\nu_3\nu_4}) \propto C_\ell^{\nu_1\nu_3}C_\ell^{\nu_2\nu_4} + C_\ell^{\nu_1\nu_4} C_\ell^{\nu_2\nu_3},
\end{equation}
where the proportionality factor includes terms dependent on $\ell$ and $f_{\textrm{sky}}$. When inverting the covariance matrix, we neglect the contribution from tensor modes in the mid-frequency channel. We finally invert the Fisher matrix to obtain the marginalised errors on each parameter \citep[see e.g.][]{Verde_2007} -- for the full expressions of these variances see Appendix \ref{ap:2}. We now take the limit $N_\ell^{M} \ll C_\ell^{dd}$, as is justified for an experiment such as SO, where the noise in the mid-frequency channel is at most \SI{1}{\percent} of the dust power spectrum in the high-frequency channel at $30 \lesssim \ell\lesssim 150$. The expressions for the variance per $\ell$ simplify to:
\begin{gather}\label{eq:sigma1}
     \sigma_\ell^2(r) \propto \frac{2 C_\ell^{dd} \left( f^2 N_\ell^H\right)^2   \left(C_\ell^{dd} + 2N_\ell^H \right)}{\left[ C_\ell^{BB} \left( C_\ell^{dd} + N_\ell^H\right) \right]^2}, \quad N_\ell^{M}/C_\ell^{dd} \ll 1,\\
      \sigma_\ell^2(f) \propto  \frac{f^2 N_\ell^H}{C_\ell^{dd}}, \quad N_\ell^{M}/C_\ell^{dd} \ll 1,\label{eq:sigma2}
\end{gather}
where the proportionality factor is $1/\left[f_{\textrm{sky}}\left(2\ell+1\right)\right]$, as there are $f_{\textrm{sky}}\left(2\ell+1\right)$ modes per multiple $\ell$. In the further limit $N_\ell^{H} / C_\ell^{dd} \ll 1$\footnote{The noise in the high-frequency channel of a SO-like experiment will be at most $\SI{10}{\percent}$ of the dust power-spectrum signal at $30 \lesssim \ell \lesssim 150$.}, the expression for $\sigma(r)$ becomes:
\begin{gather}\label{eq:sigma11}
    \sigma_\ell^2(r)= \frac{2f^4}{f_{\textrm{sky}}\left(2\ell+1\right)} \Bigg[ \frac{ N_\ell^H }{C_\ell^{BB}} \Bigg]^2 + \mathcal{O}((N_\ell^H /C_\ell^{dd})^2).
\end{gather}
A direct evaluation of this expression for a SO-like configuration ($f_{\textrm{sky}} = 0.1$, $H$ referring to the \SI{225}{\giga\hertz} ultra-high-frequency channel) gives $\sigma(r) = \SI{1.3e-3}{}$ at $\ell = 80$, in agreement with the value found in our baseline MCMC analysis. Moreover, we explicitly see from equation \eqref{eq:sigma11} that \sigmar is independent of the dust power spectrum up to second order in $N_\ell^H/C_\ell^{dd}$. Moreover, as $N_\ell^{H}$ goes to zero, the error on $r$ gets arbitrarily small thanks to sample variance cancellation \citep{Schmittfull_2018}. This limit also results in a small error in the dust parameter $f$. Thus, we see that the error on $r$ is unaffected by the statistical properties of the dust (making only reasonable assumptions about the levels of dust and noise).

The intuition for these findings is that when measuring the power spectra in the high-frequency channels, we are measuring the actual dust fluctuations. This means that one can simply subtract them without knowing the specific statistical realization of the dust, simply because the fluctuations are the same on both high- and mid-frequency channels (provided negligible levels of frequency decorrelation). For this to work, we need an estimate of $f$ that is not limited by our knowledge of the dust statistics. This was proven in our forecast, for which the error on $f$ tends to zero as the noise in the high-frequency channel becomes insignificant. Another way of understanding this result is that, with a good enough estimate of the dust frequency-scaling $f$, the MF channel (where the CMB signal is most prominent) can be cleaned from dust using the UHF channel (where dust is dominant) by a simple subtraction at the map level. The power-spectrum residuals in the MF map will scale as $\sigma_\ell^2(f)$ (equation \ref{eq:sigma2}) times $C_\ell^{dd}$, which is indeed independent of $C_\ell^{dd}$ itself. This map-level subtraction holds for whatever the dust realization may be (Gaussian or not). As a result, \sigmar is not affected. In fact, this technique has been implicitly used in most map-based component separation methods \citep{deBelsunce_2022,deBelsunce_2023, Efstathiou_2009, Efstathiou_2019}. We have now explicitly shown why these methods are immune to  the statistics of the foreground fields; given this result, it appears that this operation at the map level is being ``implicitly'' performed in the $C_\ell$-likelihood. For the toy model at hand, where all parameters are linear, one can show that the power-spectrum likelihood indeed yields the same results as a map-based likelihood. Although $r$ is unaffected, from equation \eqref{eq:sigma2} we can see that our ability to constrain the dust field properties is directly determined by our measurement of the dust field itself. Hence, the absence of a reliable method to quantify the level of foreground non-Gaussianity directly impacts the error estimation on the spatial dust parameters $A_d$ and $\alpha_d$.

\subsection{Impact on goodness-of-fit}\label{sec:3-pte}

We have seen that $C_\ell$-based foreground cleaning constraints on \rtensor~are not impacted by a non-Gaussian, anisotropic dust field.  However,  we see that the uncertainties on the dust spatial parameters ($A_d, \alpha_d$) can become significantly underestimated in the Gaussian approximation. We now investigate how this impacts goodness-of-fit $\chi^2$ values, which are common way to assess the suitability of the model being used in the likelihood \citep[e.g. \citetalias{BICEP_2021};][]{Wolz_2023} or to search for systematic errors in the measurements.

\begin{figure}
     \centering
     \includegraphics[width =0.9\columnwidth]{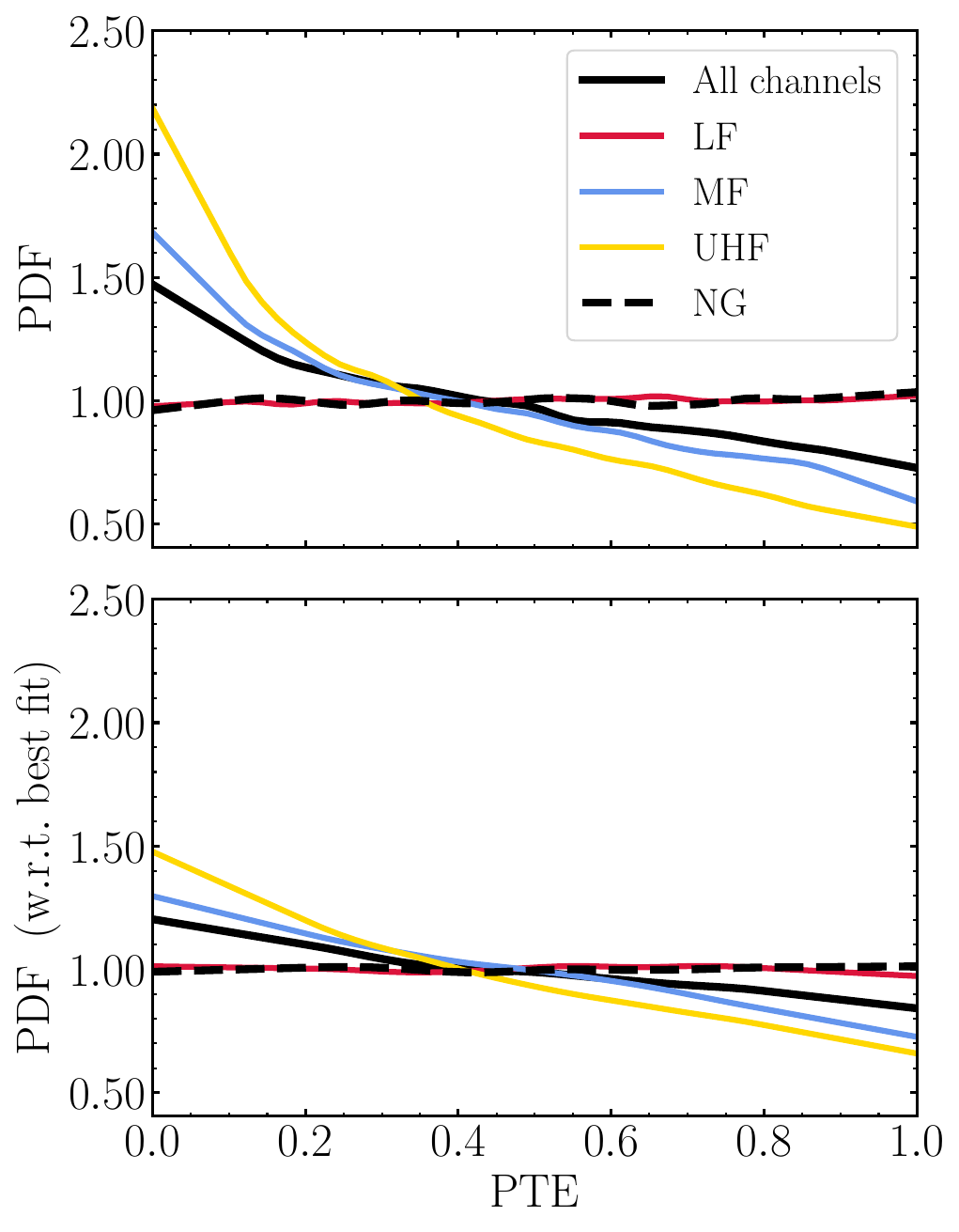}
     \caption{Histogram of PTE values obtained for \SI[print-unity-mantissa=false]{1e5} power-spectrum mock simulations drawn from \covng, then analysed with \covng~(dashed line) and with \covg~(solid line). In the top panel, the PTE is obtained directly from the raw simulations with respect to the true model. In the bottom panel, we first perform a best fit against which we compare the mock data. Different colours correspond to the inclusion of different channels. With the exception of the low-frequency (LF) channel, which barely contains any dust, the PTE distributions are skewed towards low values. This would be interpreted as the model being unable to provide a good description of the data, whereas the actual problem is a mis-estimation of the data vector uncertainties.}
     \label{fig:pte_comparison}
 \end{figure}

We take the \SI[print-unity-mantissa=false]{1e5} mock power-spectrum observations $\hat{C}_\ell$ generated by performing a multivariate Gaussian sampling of \covng. We study two cases. In the first case, we directly compute the $\chi^2$ value, defined as:
\begin{equation}\label{eq:chi2}
    \chi^2 =  (\hat{C}_\ell - C_\ell^{\rm{model}})^{\rm{T}}  \Sigma^{-1}~(\hat{C}_\ell - C_\ell^{\rm{model}}),
\end{equation}
where $C_\ell^{\rm{model}}$ refers to our initial data vector (Section \ref{sec:2-mock}) that we generated with the sky model presented in Section \ref{sec:2-skymodel}. The $\chi^2$ value is computed twice, for both $\Sigma=$~\covg~and $\Sigma=$~\covng.  In the second case, we first refit the mock observations and obtain best-fitting values ${C}^{\textrm{bf}}_\ell$. The $\chi^2$ value is then computed for the mock data vector with respect to (w.r.t.) the best-fitting model:
\begin{equation}\label{eq:chi2bestfit}
    \chi^2_{\textrm{bf}} =  (\hat{C}_\ell - C_\ell^{\rm{bf}})^{\rm{T}}  \Sigma^{-1}~(\hat{C}_\ell - C_\ell^{\rm{bf}}),
\end{equation}
for both $\Sigma = $\covg, \covng. Finally, we obtain the probability to exceed (PTE) of the observed $\chi^2$ value for each simulation. In order to convert from a $\chi^2$ value to a PTE, we assume the numbers of degrees of freedom (d.o.f.) to be the same as the number of data points in the first case, and the number of data points minus the number of model parameters in the second case. 

We show the PTE histograms in the form of a probability density function (PDF) in Figure \ref{fig:pte_comparison}. The top panel corresponds to the first case (equation \ref{eq:chi2}), and the bottom panel corresponds to the second case (equation \ref{eq:chi2bestfit}, where the $\chi^2$ is calculated with respect to each best-fitting model). The dashed line represents the analysis using \covng, for which we obtain a uniform distribution since the $\chi^2$ is computed with the same covariance that generated the data. On the other hand, the solid black line represents the PTEs obtained with \covg. This PTE distribution is skewed towards low values, meaning that there is a substantial number of simulations where the $\chi^2$ are very high and significantly differ from the expected value for the data: \SI{8}{\percent} of simulations have $\textrm{PTE} < 0.05$ (\SI{6}{\percent} if compared to the best fit) as opposed to the expected $\SI{5}{\percent}$ value. This would be interpreted as the model being unable to provide a good description of the data, whereas the actual problem is a mis-estimation of the data vector uncertainties.

Similar results were found in \cite{Azzoni_2023}, where the authors assumed a Gaussian covariance when analysing simulated data in which the dust component, generated using the model of \cite{Vansyngel_2017}, was significantly non-Gaussian. They found that, while the theoretical model used to describe the power spectra was able to obtain unbiased constraints on $r$, the best-fitting $\chi^2$ values were consistently high.

We further calculate the PTE distributions corresponding to specific frequency channels (LF, MF, UHF). When comparing to the best-fitting model, the d.o.f. to use is unclear a priori because the best fit was found including all frequency channels. We therefore resort to the distributions obtained with \covng. For each band, we find the effective d.o.f. that produces a uniform distribution. This is then used for the equivalent calculation using \covg. The results are included as differently colour-coded solid lines in Figure \ref{fig:pte_comparison}. We are only able to recover the true distribution for the LF channels. This is because they barely contain any dust. As a result, a non-Gaussian dust covariance has a negligible effect on them. The strongest effect is seen for the UHF channels, which are dust dominated. Even though comparing to the best fit alleviates the differences, neglecting the impact of dust non-Gaussianity in the covariance matrix still leads to poor goodness-of-fit metrics.
\begin{figure}
     \centering
     \includegraphics[width =0.94\columnwidth]{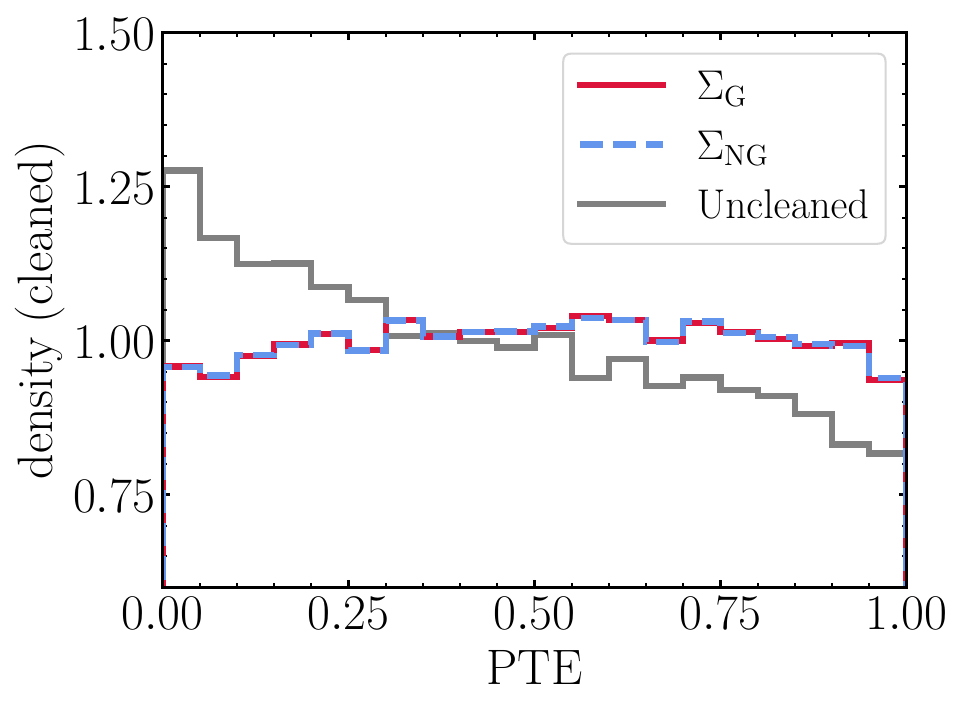}
     \caption{Similar to the bottom panel of Figure \ref{fig:pte_comparison}, but we now show the PTE distributions obtained after cleaning the \SI[print-unity-mantissa=false]{1e5}  power-spectrum mock simulations with the procedure outlined in Section \ref{sec:3-clean}. The cleaning procedure nulls the foregrounds and therefore minimizes the effect of their non-Gaussian statistics. We recover a uniform distribution for \covg, which reflects how it correctly summarises the statistics of the data after cleaning. This contrasts with the originally skewed PTE distribution found for \covg ~(solid black line, bottom panel Figure \ref{fig:pte_comparison}), added here as a solid gray line for comparison.}
     \label{fig:pte_clean}
 \end{figure}

\subsubsection{Recovering good PTEs with cleaned maps} \label{sec:3-clean}
\begin{figure*}
    \centering
    \includegraphics[width = 0.7\textwidth]{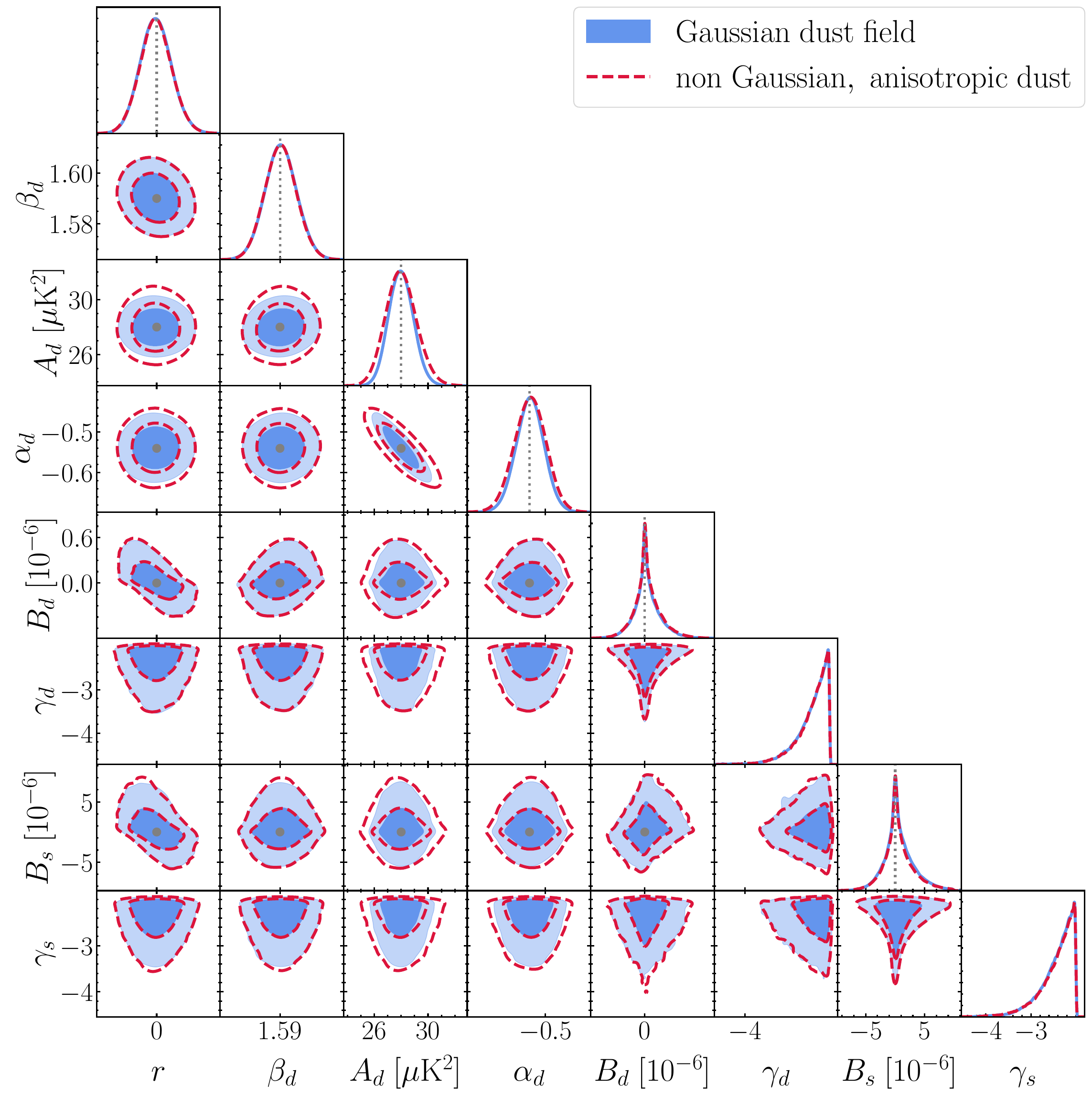} 
    \caption{Similar to Figure \ref{fig:posterior_baseline}, we show the comparison between posterior distributions obtained with a Gaussian and non-Gaussian dust covariance matrix (shown in filled solid blue and unfilled dashed red, respectively). In this case, the data have been analysed with the minimal moments expansion method of \protect\citet{Azzoni_2021} to assess the impact of frequency decorrelation. This adds 4 more parameters to our model (see equation \ref{eq:moments}), although we only show a reduced dataset for clarity. The amplitude of the spectral index power-spectrum fluctuations ($B_d$, $B_s$) are consistent with zero. As a result, we ignore the prior volume effects that appear on the corresponding $\gamma_d$, $\gamma_s$. We also find the same results as our baseline analysis: we recover the same unbiased parameter constraints in both (\covg, \covng) cases, with the exception of the dust spatial parameters ($A_d$, $\alpha_d$), whose variance increases. Due to the complexity of this model, the error on $r$ increases by more than \SI{50}{\percent}, $\sigma(r) = \SI{2.0e-3}{}$, with respect to the baseline analysis.}   \label{fig:posterior_moments}
\end{figure*}

The reason for the degradation in the $\chi^2$ PTE described in the previous section is our inability to fully describe the statistical uncertainties of the dust power spectrum. However, as we have shown in Section \ref{sec:3-baseline}, and justified in Section \ref{sec:3-fisher}, this mostly affects the final constraints on parameters related to the dust power spectrum, and does not impact the cosmological and spectral parameters. It should therefore be possible to select a sector of the data vector where the effects of dust non-Gaussianity are minimized, and for which standard goodness-of-fit metrics can be used as a reliable diagnostic of model suitability. This can be achieved by extracting a ``cleaned'' CMB power spectrum. A similar approach at the map-based level was developed by \cite{deBelsunce_2022} and applied to \planck~polarization maps.

Consider the harmonic coefficients of maps observed at $\mathcal{N}_\nu$ different frequencies, and let us model them as arising from a linear combination of components with spectra $S_c^\nu$ and instrumental noise:
\begin{equation}
    m_{\nu, \ell m} = \sum_{c = 1}^{\mathcal{N}_\nu} S_{c}^{\nu} A^{c}_{\ell m } + n_{\nu, \ell m},
\end{equation}
where $A^c_{\ell m}$ are the harmonic coefficients of component $c$, and $n_{\nu, \ell m}$ is the noise contribution. Assuming we know the spectra of all components, the maximum-likelihood estimator for the component amplitudes is
\begin{equation}
   \vec{A}_{\ell m} = \left({\sf S}^\intercal {\sf N}_\ell^{-1} {\sf S}\right)^{-1} {\sf S}^\intercal {\sf N}_\ell^{-1}  \vec{m}_{\ell m},
\end{equation} 
where the matrix ${\sf S}$, of size $\mathcal{N}_c \times \mathcal{N}_\nu$, contains the SEDs of all components, ${\sf N}_\ell$ is the ${\mathcal{N}}_\nu \times {\mathcal{N}}_\nu$ noise power spectrum matrix, and we have promoted $m_\nu$ and $A^c$ to vectors of sizes $\mathcal{N}_\nu$ and $\mathcal{N}_c$ respectively. If we are only interested in the CMB amplitude, we can express it as a linear combination of the input maps:
\begin{equation}
    A^{\textrm{CMB}}_{\ell m} = \sum_\nu Q_{\nu,\ell} m_{\nu, \ell m},
\end{equation}
where
\begin{equation}
    Q_{\nu,\ell} \equiv \sum_c \left({\sf S}^\intercal {\sf N}_\ell^{-1} {\sf S}\right)^{-1}_{\textrm{CMB},c} \left({\sf S}^\intercal {\sf N}_\ell^{-1} \right)^c_{\nu}.
\end{equation}
Then, the power spectrum of the cleaned CMB map is
\begin{equation}\label{eq:cleancell}
    C_\ell^{\textrm{CMB}} = \sum_{\nu,\nu^\prime} Q_{\nu,\ell} Q_{\nu^\prime,\ell} C_\ell^{\nu \nu^\prime},
\end{equation}
and its covariance is related to the full multi-frequency covariance via
\begin{align}\nonumber
    \textrm{Cov}(C_\ell^{\textrm{CMB}}, C_{\ell^\prime}^{\textrm{CMB}}) = \sum_{\substack{\nu_1,\nu_2,\\\nu_3,\nu_4}} & Q_{\nu_1,\ell}Q_{\nu_2,\ell}Q_{\nu_3,\ell'}Q_{\nu_4,\ell'} \\
    &\times\textrm{Cov}(C_\ell^{\nu_1\nu_2}, C_{\ell^\prime}^{\nu_3\nu_4} ),\label{eq:cleancov}
\end{align}
which is the same expression derived in \cite{Azzoni_2023} within the context of residual covariances in component separation methods.

We take our \SI[print-unity-mantissa=false]{1e5} mock power-spectrum observations. With their best-fitting values for the SED parameters, we compute the elements $Q_\nu$, specific to the SO frequency channels and noise spectra. We then clean the mock observations and the best-fitting signal estimates ${C}^{\rm{bf}}_\ell$ with equation \eqref{eq:cleancell}, and obtain the clean covariance matrix (equation \ref{eq:cleancov}). This is performed twice, once for the best fit and covariance derived from \covg~and once for the corresponding quantities using \covng. We compute the corresponding $\chi^2_{\textrm{bf}}$ values for each cleaned mock observation. In order to convert these to PTE values, we proceed as in the previous section and compute the effective number of d.o.f. that return a uniform PTE distribution in the \covng~case, which is the non-Gaussian covariance from which the mock observations were generated.

We show the corresponding PTE distributions in Figure \ref{fig:pte_clean}. The distribution we recover for \covg~(solid red line) exactly matches that of \covng~(blue dashed line), in contrast with the previous disagreement found when no cleaning was performed (Section \ref{sec:3-pte}) and which we overplot as a solid gray line for comparison. This highlights the suitability of the procedure to null the non-Gaussian foregrounds. It can therefore be used to quantify the ability of the model to describe the data when considering only the sources whose statistical distribution we have more under control (the CMB, in this case). It is another goodness-of-fit metric that should be explored in addition to the standard $\chi^2$, to ward off issues with non-Gaussianities.

\subsection{Foreground extensions}\label{sec:3-extra}

Perhaps the most important challenge for component separation is foreground frequency decorrelation, which may arise from inaccurate modelling of the foreground SEDs, as well as through the spatial variation of these SEDs. This has been noted in the context of map-based techniques \citep[e.g.][]{Armitage_2012, Kogut_2016, McBride_2023}, as well as in $C_\ell$-based methods \citep{BK14, Planck_2017_spatial}. It can be addressed by extending the foreground model used in component separation, to directly account for these spatial variations at the map level \citep{Eriksen_2004, Eriksen_2008, Alonso_2017, Josquin_2019} or by forward-modelling the impact of this contamination on to the multi-frequency power spectra, either through a moment expansion \citep{Chluba_2017, Rotti_2021, Remazeilles_2021, Mangilli_2021,  Azzoni_2021,Vacher_2022, Sponseller_2022, Vacher_2023b}, or through an effective description of frequency decorrelation \citep{BK14}. We here investigate whether the presence of decorrelation would affect our results, since the qualitative explanation in Section \ref{sec:3-fisher} relies on the assumption of \SI{100}{\percent} correlation between frequency channels.

To do this, we first repeat our analysis using an extended foreground model, through the so-called minimal moment expansion method of \citet{Azzoni_2021}. The model allows for spatially-varying spectral indices for both dust and synchtrotron, and parametrizes those variations as uncorrelated Gaussian random fields with power-law-like power spectra. Since we are only interested in the changes that foreground non-Gaussianities introduce at the level of the power-spectrum covariance, we analyse data without any spectral index variations. This is equivalent to evaluating the parameters of the minimal moment expansion model at $B_c, B_d, \gamma_d, \gamma_c = 0$. The results obtained using this model for component separation are presented in Figure \ref{fig:posterior_moments}. We find the same results as our baseline analysis: we recover the same unbiased parameter constraints in both (\covg, \covng) cases, with the exception of the dust spatial parameters $\alpha_d$ and $A_d$ whose variance increases a factor $1.2$ and $1.3$, respectively. We obtain null amplitudes for the spectral index power-spectrum fluctuations ($B_s$, $B_d$). Due to the complexity of this model\footnote{We are now marginalising over a model that allows for frequency decorrelation, and thus expect the additional uncertainty to leak into $r$.}, the error on $r$ increases in both cases by more than \SI{50}{\percent}, $\sigma(r) = \SI{2.0e-3}{}$, with respect to the baseline analysis where no moments method was used. This agrees with the increase found in \cite{Azzoni_2021}.

The results obtained so far have been recovered from a data vector which itself contained no frequency decorrelation (even if the moment expansion model allows for it). To test if the presence of significant decorrelation in the data would change our results, we generate mock data that include decorrelation via the scale-independent decorrelation parameter $\Delta_d$ presented in Section \ref{sec:2-skymodel}. It is worth noting that the moment expansion model is a generalisation of this simple constant-decorrelation model, the latter corresponding to the limit in which spectral index variations are uncorrelated between pixels \citep{Vansyngel_2017, Azzoni_2023}. The amount of decorrelation present in data is still highly disputed. \cite{Sheehy_2018} found negligible levels of decorrelation between the \SI{217} and \SI{353}{\giga\hertz} \planck~channels, a result also found independently in \cite{Planck_2020_dust}. No significant evidence of decorrelation was found in the in the BICEP/Keck region either \citepalias{BICEP_2021}, a result that was further supported in an analysis including correlations with neutral hydrogen \citep{BICEP_2023_HI}. Nevertheless, decorrelation between the foreground emission spectral laws at different frequencies can be detected on individual lines of sight, as noted by \cite{Pelgrims_2021}, and hence it is worth exploring the potential impact of its presence on our conclusions.

\begin{figure}
    \centering
    \includegraphics[width = 0.9\columnwidth]{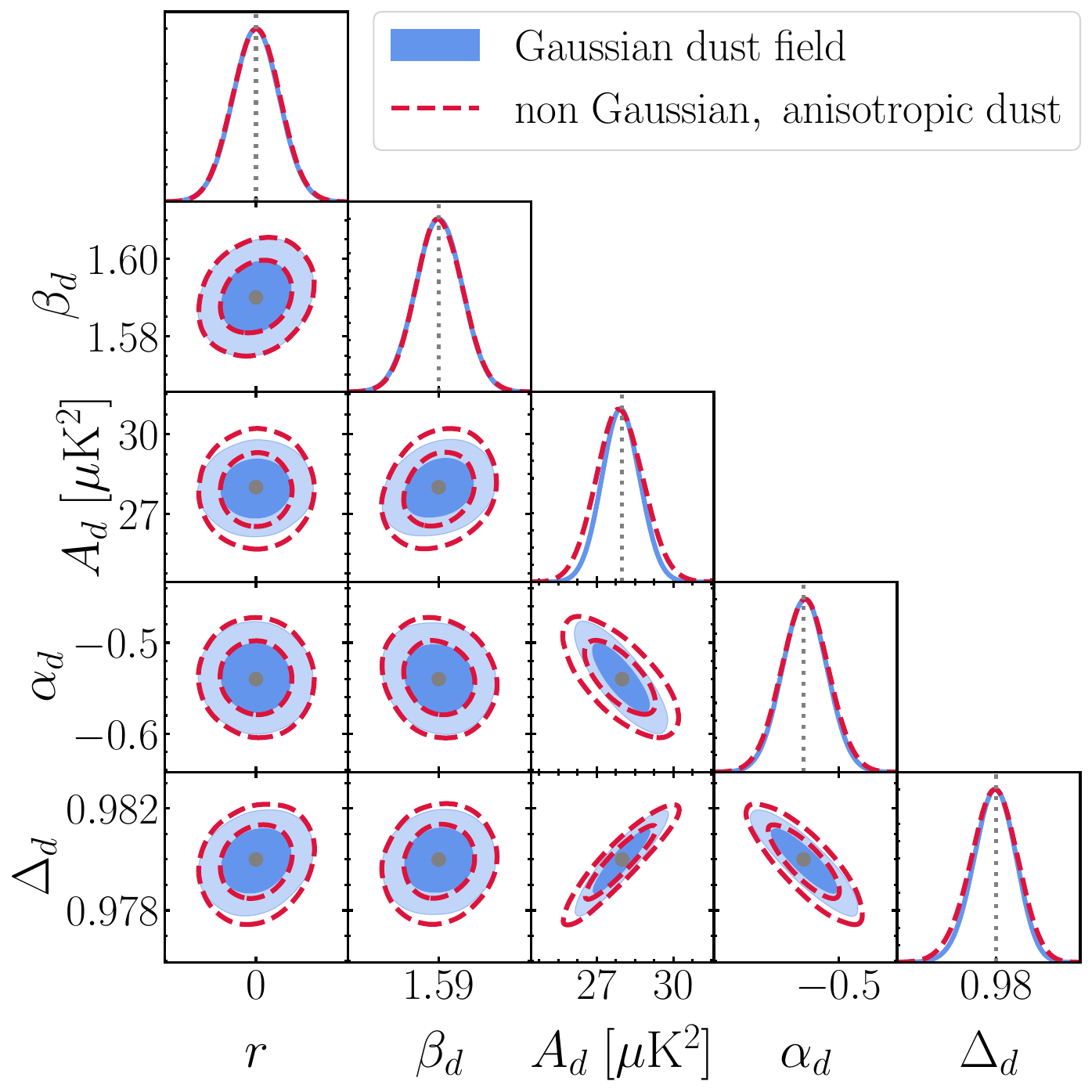} 
    \caption{Similar to Figure \ref{fig:posterior_baseline}, we show the comparison between posterior distributions obtained with a Gaussian and non-Gaussian dust covariance matrix (shown in filled solid blue and unfilled dashed red, respectively) for the most relevant parameters included in the sky model with simple $\Delta_d$ decorrelation. The scale-independent parameter $\Delta_d$ characterizes the level of decorrelation between the two UHF channels. In this case, the dust power-spectrum signal itself contains $\SI{2}{\percent}$ decorrelation, although this level is small enough not to be captured in the covariance.  We also find the same results as our baseline analysis: we recover the same unbiased parameter constraints in both (\covg, \covng) cases, with the exception of the dust spatial parameters ($A_d$, $\alpha_d$), whose variance increases. The complexity of decorrelation increases the error on \rtensor~by \SI{30}{\percent}, $\sigma(r)=\SI{1.7e-3}{}$.}  \label{fig:posterior_decorr}
\end{figure}

We choose to generate data with \SI{2}{\percent} decorrelation between the \SI{217} and \SI{353}{\giga\hertz} channels for the SO SAT region. This level of decorrelation is compatible with existing data \citep{Planck_2020_dust, Pelgrims_2021}. While the power-spectrum values change, this small level of decorrelation allows us to use the same covariance matrix as in all previous tests. This incurs a \SI{2}{\percent} error on the cross-frequency correlations and is expected to be irrelevant when computing the covariance matrix.

We run the MCMC inference with both data and model now including scale-independent decorrelation in the dust component. We show our posteriors in Figure \ref{fig:posterior_decorr}. While the uncertainty on $r$ has now increased by \SI{30}{\percent}, $\sigma(r)  = \SI{1.7e-3}{}$, we recover the same behaviour observed in the baseline analysis (Figure \ref{fig:posterior_baseline}). Firstly, we obtain unbiased constraints for all parameters. Secondly, we find that the constraint on $\sigma(r)$ is unaffected by the non-Gaussian statistics of dust at this level of decorrelation. Finally, we see a $1.1$ and $1.2$ factor increase in the errors of the spatial dust parameters $\alpha_d$ and $A_d$, respectively, when the data are analysed with \covng. Thus, as long as it is included in the modelling, this level of decorrelation does not hamper our ability to employ dust maps at one frequency to constrain the dust emission at another one.

If decorrelation is not taken into account within the model, we see an increase in $\sigma(r)$. In particular, when we analyse data containing $\Delta_d = 0.98$ with our baseline model (which does not include decorrelation),  we obtain $\sigma(r)=\SI{3.5e-3}{}$ (which corresponds to a factor of 2 increase)\footnote{In practice, however, this value for $\sigma(r)$ is not reliable. This model would never be used to analyse the dataset as the PTE goodness-of-fit metric is less than $10^{-9}$.}. This illustrates how the interpretation obtained with our simple toy model (equation \ref{eq:model}) only applies when all frequency channels are perfectly correlated. If a significant level of decorrelation is present, a potential solution would be to break up the sky into different patches, within which the spectral index is constant and the implicit map-level subtraction can be performed (provided the patches are signal dominated and the spectral fluctuations are slowly varying).

\section{Conclusions}\label{sec:4}

A detection of the primordial gravitational wave background is of paramount importance: it would provide a conclusive test of inflation and open a whole new window to probe physics at the highest energies and earliest times. However, exquisite control over the polarized foreground signal is key for a reliable detection of primordial \B-modes. In this paper, we addressed the standard assumption made in $C_\ell$-based cleaning methods, namely that foregrounds can be approximated as Gaussian \citepalias[e.g.][]{BICEP_2021}. Focusing on dust, its covariance matrix is modified to include the effects of the anisotropic and non-Gaussian nature of this field. Both effects lead to additional couplings between different $\ell$ bins, as well as modifying the power-spectrum uncertainties.

We studied two regimes of non-Gaussianity, introduced by 1) a large-scale modulating field, and 2) small-scale filamentary dust structures. In 1), the large-scale template, which we took to be the \planck~\SI{353}{\giga\hertz} temperature map smoothed on small scales, introduces couplings between neighbouring $\ell$ bins close to the dust covariance matrix diagonal. In 2), we obtained the covariance on small-scales using 200 realizations of the \mbox{\textsc{dustfilaments}} model \citep{DF_2022}. These filamentary structures couple large and small scales, and also modify the diagonal elements of the covariance matrix.

We compared the CMB and dust parameter inference obtained with a dust covariance matrix that captures these non-Gaussianities against one assuming Gaussian foregrounds. We selected noise levels and sky coverage such that we replicate the SO experiment, although we checked that our results are robust against other configurations more similar to a \litebird-like experiment. We found that the final constraints on \rtensor~are \emph{not} affected by non-Gaussian dust foregrounds, although the error on the spatial dust parameters increases by a factor $\sim 1.2$ on average. This reflects how their uncertainties can be significantly underestimated in the Gaussian approximation. We believe these results hold in general, as we recovered the same behaviour for \rtensor~when we artificially set all off-diagonal covariance matrix elements to have different, very large values.

We explained these results by arguing that $C_\ell$-based constraints on \rtensor~are nearly immune to the statistics of the dust field, as one can always clean the CMB signal from dust by a simple subtraction of the particular realization of dust on the sky using a dust-dominated high-frequency channel. We checked this conclusion using a simple analytical Fisher forecast that assumes \SI{100}{\percent} correlation between frequency channels, although we found the same behaviour numerically for data that include decorrelation using an MCMC analysis. We also expect map-based component separation methods to be immune to the statistics of the foreground fields, given that similar arguments about removing the exact realization of the foreground field, without requiring knowledge of its statistics, apply to map-based cleaning methods.

We note, however, that even though \rtensor~constraints are not affected, dust non-Gaussianity noticeably impacts goodness-of-fit analyses. This behaviour was also found by \cite{Hoz_2022} in the context of determining polarization angles in CMB experiments. The $\chi^2$ values computed with a covariance representing Gaussian foregrounds are degraded not because our model is unable to provide a good description of the data, but because we are unable to fully describe the statistical fluctuations of the dust power-spectrum through the Gaussian covariance matrix. We explored how this can be mitigated by using only foreground-cleaned spectrum combinations when computing goodness-of-fit statistics. 

In this paper, we have considered non-Gaussianity in the foreground amplitudes, but not in the spectral index variations. The latter could add additional contributions to the non-Gaussian covariance that produce a non-trivial frequency structure. Without better data to model these SED variations, it is difficult to construct a well-educated model, so we leave this task for future work.

The arguments presented in this work should also apply to non-Gaussianities present in the synchrotron background, although we leave such a study to future work. In addition, our work highlights the importance of taking into account non-Gaussianities in any kind of Galactic science study. Although the work presented here focused on the Simons Observatory, its results will be relevant for any experiment that interprets \B-mode constraints on the tensor-to-scalar ratio in the absence of a reliable method to quantify the level of foreground non-Gaussianity.

\section*{Acknowledgements}

The authors acknowledge the efforts of Susanna Azzoni and Kevin Wolz, who contributed to the development of the component separation likelihood used in this work. We thank Steven Gratton and Anthony Challinor for helpful input and discussions, and Erik Rosenberg for sharing his expertise on using \planck~data. We also thank Susanna Azzoni, Ant\'on Baleato Lizancos and Kevin Wolz for providing insightful comments on an early paper draft. IAC acknowledges support from Fundaci\'on Mauricio y Carlota Botton and the Cambridge International Trust. CHC acknowledges funding by ANID, FONDECYT Postdoc fellowship 3220255 and by ANID, BASAL, FB210003. SvH was supported by the Carlsberg Foundation, acknowledges support by the Beecroft Trust, and wishes to thank Linacre College for the award of a Junior Research Fellowship. BDS acknowledges support from the European Research Council (ERC) under the European Union’s Horizon 2020 research and innovation programme (Grant agreement No. 851274). BDS further acknowledges support from an STFC Ernest Rutherford Fellowship. DA acknowledges support from the Science and Technology Facilities Council through an Ernest Rutherford Fellowship, grant reference ST/P004474.

Some of the results in this paper have been derived using the \textsc{healpy} \citep{Zonca_2019} and \textsc{healpix} \citep{Healpix_2005} packages. This research has also made extensive use of the \textsc{namaster} \citep{namaster}, \textsc{sacc} \citep{sacc}, \textsc{astropy} \citep{astropy}, \textsc{numpy} \citep{numpy}, \textsc{scipy} \citep{scipy} and \textsc{emcee} \citep{emcee_2013} packages. We also acknowledge use of the \textsc{matplotlib} \citep{matplotlib} and \textsc{getdist} \citep{getdist} packages to produce the plots in this paper. This research also used resources of the National Energy Research Scientific Computing Center (NERSC), a U.S. Department of Energy Office of Science User Facility located at Lawrence Berkeley National Laboratory, operated under Contract No. DE-AC02-05CH11231 using NERSC award HEP-ERCAP-mp107. The Geryon cluster at the Centro de Astro-Ingenieria UC was used for some of the 
simulations performed in this paper. ANID BASAL project FB21000, BASAL CATA PFB-06, the Anillo ACT-86, FONDEQUIP AIC-57, and QUIMAL 130008 provided funding for several improvements to the Geryon cluster.

\section*{Data Availability}

The \planck~\SI{353}{\giga\hertz} map was accessed from the \planck\footnote{\planck~(\url{https://www.esa.int/Planck}) is an ESA science mission with instruments and contributions directly funded by ESA Member States, NASA, and Canada.}~Legacy Archive at \url{https: //pla.esac.esa.int/}. The small-scale dust filament maps were generated using the \DF~\citep{DF_2022} simulation. The code to generate all other data is available at \url{https://github.com/iabrilcabezas/DustNGwBBP}, and will be shared on reasonable request to the corresponding author. 


\bibliographystyle{mnras}
\DeclareRobustCommand{\DELA}[3]{#3}
\DeclareRobustCommand{\DE}[3]{#3}
\bibliography{main} 


\appendix

\section{Non-Gaussian anisotropic dust covariances} \label{sec:ap1}

\subsection{Large-scale modulating field}\label{sec:ap1-norm}

As discussed in the main text (Section \ref{sec:2-cov}), we write down a dust model where the Gaussian isotropic dust field $\delta(\nv)$ is modulated by a large-scale intensity field $t(\nv)$ \citep{Mak_2017}:
\begin{gather}
    d_{\textrm{NG}}(\nv) = w(\nv)t(\nv)\delta(\nv),
\end{gather}
where $w(\nv)$ is the sky mask. This model ensures that small-scale variance of the dust emission scales with the square of the large-scale intensity, as found empirically \citep{Miville_2007}. 

This model is compared against one where no modulation is present:
\begin{gather}
    d_{\textrm{G}}(\nv) = w(\nv)\delta(\nv).
\end{gather}
One can understand the effect of $t(\nv)$ as producing an effective sky mask $\tilde{w}(\nv)\equiv w(\nv) t(\nv)$. For both cases we can define the corresponding pseudo-$C_\ell$, whose ensemble average can be expressed as:
\begin{gather}
\langle\tilde{C}^{d_\textrm{G}}_\ell\rangle =\sum_{\ell^\prime}M^{w}_{\ell\ell^\prime}\,C_{\ell'}^\delta, \label{eq:pclg}\\
\langle\tilde{C}^{d_\textrm{NG}}_\ell\rangle =\sum_{\ell^\prime}M^{\tilde{w}}_{\ell\ell^\prime}\,C_{\ell'}^\delta,\label{eq:pclng}
\end{gather}
where $C^\delta_\ell$ is the power spectrum of $\delta(\nv)$, and we have defined the $C_\ell$-level mode-coupling matrices ${M}_{\ell\ell^\prime}^{x}$, which depend on the mask $x(\nv)$ field alone and can be computed analytically with \textsc{namaster} \citep{namaster}. One property of ${M}_{\ell\ell^\prime}^{\tilde{w}}$ is that
\begin{equation}  \label{eq:skyavw}\sum_{\ell'}M^{\tilde{w}}_{\ell\ell'}=\int\frac{d\nv}{4\pi}\tilde{w}^2(\nv)\equiv\langle \tilde{w}^2\rangle_\Omega,
\end{equation}
where $\langle\cdots\rangle_\Omega$ denotes sky averaging. If $C_\ell^\delta$ is smooth, it can be pulled out of the sums in equations (\ref{eq:pclg}, \ref{eq:pclng}) to yield:
\begin{equation}
\langle\tilde{C}^{d_\textrm{NG}}_\ell\rangle=C_\ell^\delta\langle\tilde{w}^2\rangle_\Omega.
\end{equation}
With this approximation, $\langle\tilde{w}^2\rangle_\Omega$ serves as the conversion factor between the two power spectra, without the need to invert the mode-coupling matrix \citep[see e.g.][]{Nicola_2021}.  

We want to reflect the presence of the anisotropic correlating template at the covariance level, but not at the power-spectrum level. As a result, we need to normalize the initial large-scale template $t_0(\nv)$ (which in our case we take to be the \planck~\SI{353}{\giga\hertz} map smoothed over \ang{0.4} scales) with the following operation:
\begin{equation}\label{eq:normt}
  t(\nv)=t_0(\nv)\sqrt{\frac{\langle w^2\rangle_\Omega}{\langle w^2t_0^2\rangle_\Omega}}.
\end{equation}
This then guarantees $\langle\tilde{C}^{d_{\textrm{NG}}}_\ell\rangle=\langle\tilde{C}^{d_\textrm{G}}_\ell\rangle=\langle w^2\rangle_\Omega\,C^\delta_\ell$.

\subsection{Covariance of large-scale modulated dust field}

We start by considering the harmonic coefficients of both $d_\textrm{G}$ and $d_{\textrm{NG}}$:
\begin{gather}
  {d}_{{\textrm{G}},\ell m}=\sum_{\ell'm'}W^{{w}}_{\ell m,\ell' m'}\delta_{\ell'm'},\\
  {d}_{{\textrm{NG}},\ell m}=\sum_{\ell'm'}W^{\tilde{w}}_{\ell m,\ell' m'}\delta_{\ell'm'},\label{eq:dNGcoeff}
\end{gather}
where $W^{x}_{\ell m, \ell^\prime m^\prime}$ are the map-level mode-coupling coefficients for $x = w, \tilde{w}$. They are such that if $x$ is slowing varying, then $W^{x}_{\ell m, \ell^\prime m^\prime}$ will have a narrow peak around $\ell = \ell^\prime, m = m^\prime$. They are also directly related to the $C_\ell$-level mode-coupling matrix:
\begin{equation}
M^{x}_{\ell\ell'}\equiv\frac{1}{2\ell+1}\sum_{mm'}|W^{x}_{\ell m,\ell'm'}|^2.\label{eq:MW}
\end{equation}

We calculate the covariance of $\tilde{C}_\ell^{d_{\textrm{NG}}}$ via computing the 4-point function of $d_{\textrm{NG}}(\nv)$, $ \langle |{d}_{\textrm{NG},\ell_1m_1}|^2|{d}_{\textrm{NG},\ell_2m_2}|^2\rangle$ using the decomposition \eqref{eq:dNGcoeff}. We use Wick's theorem, assuming $\delta(\nv)$ to be Gaussian, to transform the $\langle\delta_{\ell_1'm_1'}\delta_{\ell_1''m_1''}^*\delta_{\ell_2'm_2'}\delta_{\ell_2''m_2''}^*\rangle$ term into different $C^{\delta}_\ell C_{\ell^\prime}^{\delta}$ contributions.  From the 4-point function we can derive the covariance of $\tilde{C}^{d_\textrm{NG}}$, ${\rm Cov}(\tilde{C}^{d_{\textrm{NG}}}_{\ell_1},\tilde{C}^{d_{\textrm{NG}}}_{\ell_2})$, which for a given pair of $\ell$ values involves sums over all other $\ell$ and corresponding $m$ values. We use the property that for a smooth $\tilde{w}$, the mode-coupling coefficients are peaked around the diagonal and also assume that the $C_\ell^{\delta}$ coefficients are slowly-varying. As a result, we approximately pull them out of the sums in the form $\bar{C}^\delta_{\ell_{12}}=\sqrt{C^\delta_{\ell_1}C^\delta_{\ell_2}}$\footnote{Note this is an approximation, and the arithmetic mean could have also been chosen, $\bar{C}^\delta_{\ell_{12}}=\frac{C^\delta_{\ell_1}+C^\delta_{\ell_2}}{2}$ .}. This, in turn, allows us to group the remaining terms into the mode-coupling matrix definition \eqref{eq:MW}. The final covariance reads:
\begin{equation}\label{eq:covng}
    {\rm Cov}(\tilde{C}^{d_{\textrm{NG}}}_{\ell_1},\tilde{C}^{d_{\textrm{NG}}}_{\ell_2}) = \frac{2(\bar{C}^\delta_{\ell_{12}})^2}{2\ell_2+1}M^{\tilde{w}^2}_{\ell_1\ell_2},
\end{equation}
where $M^{\tilde{w}^2}$ is the mode-coupling matrix for the mask $\tilde{w}^2(\nv) = \tilde{w}(\nv)\tilde{w}(\nv)$. In the case of $d_\textrm{G}$, the derivation is analogous, with the only difference arising from the updated sky mask, which no longer contains the modulating template $t(\nv)$:
\begin{equation}
  {\rm Cov}(\tilde{C}^{d_\textrm{G}}_{\ell_1},\tilde{C}^{d_\textrm{G}}_{\ell_2}) = \frac{2(\bar{C}^\delta_{\ell_{12}})^2}{2\ell_2+1}M^{w^2}_{\ell_1\ell_2}.
\end{equation}

We generalise these expressions to the case of multi-frequency power spectra with the transformation:
\begin{equation}
2(\bar{C}^\delta_{\ell_{12}})^2\hspace{6pt}\rightarrow\hspace{6pt}\bar{C}_{\ell_{12}}^{\nu_1\nu_3}\bar{C}_{\ell_{12}}^{\nu_2\nu_4}+\bar{C}_{\ell_{12}}^{\nu_1\nu_4}\bar{C}_{\ell_{12}}^{\nu_2\nu_3}.
\end{equation}

Finally, we correct equation \eqref{eq:covng} to obtain the covariance of the dust field itself, $D(\nv) = t(\nv)\delta(\nv)$. Since the normalization we perform on $t(\nv)$ \eqref{eq:normt} ensures that $C_\ell^{\delta} = C_\ell^D$, we can also write
\begin{equation}
\langle\tilde{C}^{d_\textrm{NG}}_\ell\rangle=C_\ell^D\langle{w}^2\rangle_\Omega.
\end{equation}
With this relation, the final covariance for the modulated dust field reads:
\begin{equation}\label{eq:finalDcov}
     {\rm Cov}({C}^{D}_{\ell_1},{C}^{D}_{\ell_2}) =\frac{ {\rm Cov}(\tilde{C}^{d_{\textrm{NG}}}_{\ell_1},\tilde{C}^{d_{\textrm{NG}}}_{\ell_2})}{\langle w^2 \rangle^2_\Omega}.
\end{equation}

\subsection{Large-scale covariance for \textsc{dustfilaments} simulations} \label{sec:ap-sumcov}

As discussed in the main text, the \DF~\citep{DF_2022} simulations do not reproduce the large-scale polarization features and cannot be used to estimate the covariance matrix at low $\ell$. We now describe how to add a large-scale covariance (derived in the previous section) to the covariance computed from these simulations (which provides accurate answers at small scales). 

Our final dust map can be split into 2 components:
\begin{equation}
    m_{\rm{T}} = m_{\rm{L}} + m_{\rm{S}},
\end{equation}
where $m_{\rm{L}}$ is a fixed large-scale template and $m_S$ describes the small-scale component, obtained by filtering out the large scales in the \DF~realization. We measure the covariance arising from the small scales, ${\textrm{Cov}}(C_\ell^{SS}, C_{\ell^\prime}^{SS})$, from a set of \DF~simulations.

We can approximate the covariance of the total map as:
\begin{equation}
\begin{split}
    {\textrm{Cov}}(C_\ell^{TT}, C_{\ell^\prime}^{TT}) = ~&  \widehat{\textrm{Cov}}(C_\ell^{TT}, C_{\ell^\prime}^{TT})~- \\
    &- \widehat{\textrm{Cov}}(C_\ell^{SS}, C_{\ell^\prime}^{SS}) + {\textrm{Cov}}(C_\ell^{SS}, C_{\ell^\prime}^{SS}),
\end{split}
\end{equation}
where $\widehat{\cov}({C}^{TT}_{\ell_1},{C}^{TT}_{\ell_2}) $ is the covariance of the large-scale modulated dust field,
\begin{equation}
      \widehat{\cov}({C}^{TT}_{\ell_1},{C}^{TT}_{\ell_2}) = \frac{2\langle\bar{C}^{TT}_{\ell_{12}}\rangle^2}{2\ell_2+1}M^{\tilde{w}^2}_{\ell_1\ell_2},
\end{equation}
for which $\langle \bar{C}_{\ell_{12}}^{TT}\rangle$ is simply the analytical power-spectra used in this work (Section \ref{sec:2-mock}). This ensures that non-Gaussianities are introduced only at the level of the power-spectrum covariance. The term $\widehat{\cov}({C}^{SS}_{\ell_1},{C}^{SS}_{\ell_2})$ is the Gaussian contribution from the small scales,
\begin{equation}
      \widehat{\cov}({C}^{SS}_{\ell_1},{C}^{SS}_{\ell_2}) = \frac{2\langle\bar{C}^{SS}_{\ell_{12}}\rangle^2}{2\ell_2+1}M^{{w}^2}_{\ell_1\ell_2},
\end{equation}
for which $\langle \bar{C}_{\ell_{12}}^{SS}\rangle$ is the average power spectra, computed over simulation ensembles of $m_{S}$ maps. We only bin the resulting covariance matrices at the last step. Therefore, in order to obtain $\langle \bar{C}_{\ell_{12}}^{SS}\rangle$ at each $\ell$, we follow the method described in \cite{Nicola_2021}: we compute the power spectra at every $\ell$ with the usual \texttt{anafast} \textsc{healpix} routine \citep{Healpix_2005} on the masked maps, and then divide this quantity by $\langle w ^2 \rangle_\Omega$. We verified that the covariance matrices estimated with this method were both symmetric and positive-definite in all cases.

\section{Toy model and Fisher forecast} \label{ap:2}

In Section \ref{sec:3-fisher}, we devised a simple model \eqref{eq:model} to gain insight into the observed behaviour for the posterior distributions (Figure \ref{fig:posterior_baseline}). The model has two parameters: $r$, the amplitude of the primordial \bb~signal and $f$, the dust emission ratio between the mid- and high-frequency channels. We assume Gaussian fields and compute the covariance matrix using the Knox formula \citep{Knox_1995}. The full marginalised errors on each parameter, obtained from the Fisher forecast are:
\begin{gather} \label{eq:apsf}
    \sigma^2(f) = \frac{f^2 N_\ell^H C_\ell^{dd} +A_\ell}{\left(C_\ell^{dd}\right)^2},\\
    \sigma^2(r) = \frac{2 \left( f^2 N_\ell^H C_\ell^{dd} + A_\ell \right) \left[ f^2 N_\ell^H \left(C_\ell^{dd} + 2N_\ell^H \right)+ A_\ell \right]}{\left[ C_\ell^{BB} \left( C_\ell^{dd} + N_\ell^H\right) \right]^2},\label{eq:apsr}
\end{gather}
where $A_\ell = \left( C_\ell^{dd} + N_\ell^H\right) N_\ell^M$. Taking into account that there are $f_{\textrm{sky}}(2\ell+1)$ modes per multiple $\ell$, one can check that these expressions simplify to equations (\ref{eq:sigma1}, \ref{eq:sigma2}) in the limit $N_\ell^{M} \ll C_\ell^{dd}$. This approximation holds for an experiment such as SO, where the noise in the mid-frequency channel is at most \SI{10}{\percent} of the dust power spectrum in the high-frequency channel.


\bsp	
\label{lastpage}
\end{document}